\documentclass{aa}  

\usepackage{graphicx}
\usepackage{txfonts}
\usepackage{lipsum}
\usepackage{subcaption}         
\usepackage{lscape}             
\usepackage{placeins}
\usepackage{hyperref}
                          
\newcommand{\te}{{\it TESS}}
\newcommand{\kms}{km\,s$^{-1}$}

\begin{document}

   \title{A family of binaries with an extreme mass ratio}

   \author{Ya\"el Naz\'e\inst{1}\fnmsep\thanks{F.R.S.-FNRS Senior Research Associate}
     \and Gregor Rauw\inst{1}
     \and Piotr A. Ko{\l}aczek-Szyma\'nski \inst{1,2}
     \and Nikolay Britavskiy\inst{3}
     \and Jonathan Labadie-Bartz\inst{4,5}
        }

   \institute{Groupe d'Astrophysique des Hautes Energies, STAR, Universit\'e de Li\`ege, B5c, All\'ee du 6 Ao\^ut 19c, B-4000 Sart Tilman, Li\`ege, Belgium
              \email{ynaze@uliege.be}
   \and
            Astronomical Institute, Department of Physics and Astronomy, University of Wroc{\l}aw, Kopernika 11, 51-622 Wroc{\l}aw, Poland
   \and Royal Observatory of Belgium, Avenue Circulaire/Ringlaan 3, B-1180 Brussels, Belgium
   \and LIRA, Observatoire de Paris, Universit\'e PSL, CNRS, Sorbonne Universit\'e, Universit\'e Paris Cit\'e, CY Cergy Paris Universit\'e, 92190 Meudon, France
   \and DTU Space, Technical University of Denmark, Elektrovej 327, 2800 Kgs., Lyngby, Denmark
   }

  \abstract
   {Multiplicity is ubiquitous among massive stars. While the stellar components usually display similar masses, some binaries with extremely low mass ratios were also observed. Some of them are primordial, while others arise from binary interactions. The identification of systems with extreme mass ratios brings valuable information, notably on the origin of fast rotation in massive stars.}
   {We identify new short-period systems with extreme mass ratios through the detection of eclipses and reflection effects.} 
   {The physical properties of a dozen newly identified cases were precisely evaluated through high-quality photometry and spectroscopy.}
   {In addition to characterizing these binaries, we found a clear signature of apsidal motion in one system, and three other systems display long-term shifts in eclipse times. }
   {All systems we reported here are composed of a massive star and a cool low-mass companion. They are therefore primordial cases. This doubles the known number of these systems in the Galaxy. In this context, it is important to note that most massive stars in these systems, as well as in previous systems reported in the literature, rotate fast (supersynchronous compared to the orbital motion). The high incidence of fast rotation in these nascent binaries provides strong constraints for star formation models.}


   \maketitle

\section{Introduction}
Massive stars are well known to display a higher multiplicity than lower-mass stars (e.g., \citealt{moe17, off23}). In addition, the components of a massive binary more often display similar masses (mass ratio $q=M_2/M_1\sim0.5-1$, with $M_1$ the mass of the most massive primary star, and $M_2$ the mass of the least massive secondary star). Nevertheless, systems with extremely low mass ratios, $q<0.2$, do exist. They are typically expected in two cases: (1) They are born as such, that is, they simply represent the tail of the mass ratio distribution. (2) They evolve into a system with an extremely low mass ratio, that is, they are the products of mass-transfer events.

Both types of systems have been observed. \citet{moe15} reported on the discovery of short-period binaries in the Large Magellanic Cloud (LMC) with extreme mass ratios (16 cases with $q<0.2$). Although radial velocity data were missing, the authors considered the low-mass secondaries to be cool pre-main-sequence (PMS) stars and called the systems ``nascent eclipsing binaries with extreme mass ratios''. Several Galactic objects have been found to belong to this category \citep{jer13,jer15,jer21,ste20,joh21,sta21,naz23,pig24}. As these Galactic studies combined spectroscopy and photometry, it was demonstrated beyond doubt that the low-mass secondaries were also the coolest stars in the systems, which confirmed their PMS status. The typical PMS signature for the Ca triplet (at 8498, 8542, and 8662\AA) was even detected in HD\,25631 thanks to a large high-quality spectroscopic dataset \citep{naz23}.

Massive post-interaction systems with low-mass secondaries have also been identified. In this case, the secondary star was initially the most massive object of the binary. It then lost its envelope in a mass-transfer event that left behind a hot low-mass stripped star. Several examples were identified in the UV range because the contrast at these wavelengths is better \citep{wan21,goe23}. Interferometric and spectroscopic data were also used to characterize them (e.g., \citealt{mou15,kle24}). Most of these stripped stars were found on long-period ($P\sim100$\,d) orbits around Be stars, but at least one star is known in the Galaxy to form a short-period pair with an O star (HD\,96670; \citealt{naz25}). In this context, it may be noted that systems in intermediate stages (i.e., with an ongoing interaction or just after it came to an end) were also detected (e.g., \citealt{har15,she20}; for a review on Be binaries, see \citealt{lab25}).

While systems with these extreme mass ratios may be rather rare, they are of the utmost importance, and it is essential to assess their evolutionary stage. For example, they can provide information on the origin of the rotation of massive stars. The primordial binaries inform us on the initial distribution of rotational properties, whereas the post-interaction cases reveal the distribution changes caused by angular momentum transfer. While the tail of massive fast rotators is often considered to be the product of binary interactions (e.g., \citealt{dem13}), the detection of primordial fast rotators indicates that at least some of them are born as such \citep{bri24}.

Observationally, both types of systems display a large contrast between the binary components, in mass and in effective temperature (hence, also in brightness). This usually makes the secondaries difficult to detect. Only the lines from the massive primary star are usually detected in the spectra, and the Doppler shifts caused by the orbital motion of the primary have a low amplitude because the companion mass is low. Large differences in the effective temperature might lead to strong reflection effects when periods are short, however, which might provide a way to identify them in photometric light curves \citep{pig24}. We identify in this paper a new set of systems with these effects. All cases also display eclipses, which allows us to lift the uncertainty caused by inclination and to further determine the true characteristics of the stars. Section 2 provides the information on the sample creation and on the data we used in the analysis. Section 3 details the results, and Section 4 discusses their interpretation.
   
\section{Sample and data}
\subsection{Sample}
Since 2018, the Transiting Exoplanet Survey Satellite (\te) mission \citep{ric15} provides high-quality photometry for nearly the whole sky. This makes it the largest provider of space-based light curves. At the start of this research work, several catalogs of eclipsing binaries using \te\ data were already available \citep{ijs21,prs22}. We extracted from them the list of stars with spectral type O or B (down to B3). In parallel, we also examined the objects that passed the so-called \te\ threshold crossing event (TCE)\footnote{https://archive.stsci.edu/tess/bulk\_downloads/bulk\_downloads\_tce.html}, i.e. objects identified as displaying ``transits'' in the automatic processing of \te\ light curves, up to Sector 71. Using Simbad\footnote{https://simbad.u-strasbg.fr/simbad/}, we identified the TCE cases associated with stars with spectral types earlier than B3 (inclusive). Finally, we compiled a list of OB stars from the general catalogs of \citet{mai13}, \citet{ree03}, and \citet{ski09}. Stars in these catalogs with a {\it Gaia} magnitude $R_p>13$\,mag were discarded, as were sources in crowded regions (i.e., those not providing at least three quarters of the total flux in the $R_p$ band of all {\it Gaia} DR3 sources within one arcminute around the target). These three lists were merged, and the \te\ light curves were then extracted and examined for each star one by one. Our sample was built from cases with short periods ($P<10$\,d) and clearly pronounced reflection effects in addition to eclipses. Cases that were already known and analyzed in the literature were discarded. 

After we identified the targets, we retrieved their $V$ magnitudes from Simbad and photogeometric distances from \citet{bai21}. For all stars, the parallaxes appear to be reliable. Their values in the {\it Gaia} DR3 catalog are higher than five times the parallax uncertainties ($RPlx>5$) and the renormalized unit weight error (RUWE) parameter is lower than 1.5. The reddenings were estimated from dust extinction maps \citep{lal22}. The online tool G-TOMO\footnote{https://explore-platform.eu/sda/g-tomo} provides $A_V$ extinctions at the positions (i.e., $RA$, $DEC$ coordinates and distances from the {\it Gaia} catalog) of the targets, which were then converted into reddenings $E(B-V)$ using the extinction factor $R_V=3.1$. At first, the bolometric corrections were evaluated using the Simbad spectral types and a standard calibration\footnote{https://www.pas.rochester.edu/$\sim$emamajek/EEM\_dwarf\_UBVIJHK \_colors\_Teff.txt}. Initial $L_{\rm BOL}$ estimates were then produced and compared to those expected for the Simbad spectral types in the same calibration. Adjustments of the spectral types could then start because they appear clearly too late or too early in some cases. Further refinement was performed considering the effective temperatures derived from the average spectra (see below) or as available in the literature. The bolometric corrections for these temperatures were estimated from the formula (Eq. 6 and Table D.13) of \citet{ped20}. The final spectral types and bolometric luminosities are provided in Table \ref{targetlist}. This table also notes the references reporting the eclipsing nature of the targets. 

Some additional information from the literature must be mentioned. Radial velocity (RV) variations were reported for V1061\,Cen by \citet{fea57}. AN\,Dor was reported as a SB2-type binary by \citet{geb22}, and \citet{sou22} further found an inclination $i=80.8^{\circ}$, an eccentricity $e=0.054$, and a periastron argument $\omega=322^{\circ}$. \citet{tka24} characterized HD\,350685 as a SB1 with $e=0$ and a velocity amplitude $K_1=34.1$\,\kms, and HD\,309317 as SB2 with $e=0$, $K_1=34.5$\,\kms\ and $K_2=78.7$\,\kms. LS\,I\,+61\,275 was mentioned as type SB1 by \citet{hua06b}, but the authors did not derive an orbital solution because only two spectra were available to them. We discuss these cases in more detail below. None of our targets belongs to a cluster, except for LS\,I\,+61\,275, which is part of IC\,1805 (whose age was estimated to be 3.5\,Myr by \citealt{sun17}).

One star was added to our sample, that is, the binary V1208\,Sco, which belongs to the NGC\,6231 cluster (3.5--5\,Myr, \citealt{ban22}). It has long been known to be an eclipsing and spectroscopic binary \citep{lev83,bal85}. Only spectroscopic data were analyzed in detail, however. The most recent SB1 solution was provided by \citet{ban22}, and the SB2 nature was reported by \citet{ban23}. No photometric modeling was done, and we therefore decided to include it here. The extinction of this system could not be derived from the dust maps, and we therefore evaluated it by comparing the observed and expected $B-V$ color for the reported spectral type (B0.5V).

Finally, it is important to note that two stars were excluded from our sample: LS\,V\,+20\,28 and HD\,298425. The first was reported to be eclipsing by \citet{ijs24}, but the {\it Kepler} light curves clearly revealed that the eclipsing object is the neighboring star LS\,V\,+20\,27 (\citealt{lac15} even further suggested that the eclipsing system is in fact a very close neighbor of LS\,V\,+20\,27, not LS\,V\,+20\,27 itself). It only appears to be eclipsing in \te\ data because the two objects are close (32\arcsec) and the light from LS\,V\,+20\,27 contaminates the \te\ data of LS\,V\,+20\,28 (but not the {\it Kepler} data because the {\it Kepler} PSF is better than that of \te). {\it Gaia} data indicated for the second star that the eclipsing binary is a neighbor at 19.5\arcsec\ of HD\,298425 \citep{mow23}. Both cases can thus be securely discarded from the category of eclipsing binaries.

\subsection{Photometry}

The \te\ satellite observes sectors of $24^{\circ}\times 96^{\circ}$ on the sky for about one month before it shifts to a neighboring region. All our targets were observed in several such sectors (see the last column of Table \ref{targetlist}), but not always with the same mode. When they were preselected, the targets were observed with a 2\,min cadence (sectors marked with an asterisk in this table), while other targets were imaged with a cadence of 30\,min in Sectors 1--26, 10\,min for Sectors 27--55, and 200\,s since Sector 56.

A pipeline that was specifically designed for \te\ data (and was inspired by the pipeline designed for the {\it Kepler} mission) performed the main reduction steps (pixel-level calibration, background subtraction, flat-fielding, and bias subtraction). For data with a 2\,min cadence (i.e., preselected targets), the MAST portal\footnote{https://mast.stsci.edu} also provides so-called conditioned light curves (PDCSAP) in addition to the simple aperture photometry (SAP) time series. These PDCSAP curves are further corrected for crowding, the limited size of the aperture, and instrumental systematics. For each target, we kept only the best-quality (quality flag $=0$) data points, and we compared the two datasets. They were often indistinguishable, or the PDCSAP curves appeared to be better (fewer long-term trends). 

For the cases that were not preselected, individual light curves were derived from image cutouts of 51$\times$51 pixels using aperture photometry performed with the Python package Lightkurve\footnote{https://lightkurve.github.io/lightkurve/}. Only data with a high quality were selected (quality\_bitmask=`hard' in task {\sc search.tesscut}). The source mask was defined from pixels with fluxes above a few times the median absolute deviation over the median flux (the exact threshold value was adapted for each target, depending on crowding). The background mask was defined by pixels with fluxes below the median flux (i.e., below the null threshold). To estimate the background, we used two methods. First, a principal component analysis, with five components, and second, a simple median. In each case, the two background-subtracted light curves were compared. Most of the time, the light curves that were corrected with the principal component analysis were similar to or better than the median-corrected ones (fewer long-term trends) and were therefore kept.

The \te\ fluxes were then converted into magnitudes, and the average magnitude was subtracted from each individual light curve. A sample of individual light curves are shown in Fig. \ref{LC} and figures showing all light curves are available at Zenodo\footnote{https://doi.org/10.5281/zenodo.17257979}. As the \te\ pixels are large (21\arcsec), the time series might be exposed to crowding issues. We examined the close (within 1\arcmin) environment of our targets in the {\it Gaia} DR3 catalog. There are no bright ($\Delta G < 2.5$\,mag) neighbors in that area, except for HD\,154407C\footnote{This is confirmed by the ``CROWDSAP'' keyword of the 2\,min light curves, which measure the contamination by neighbors and is below 0.85 for all targets observed at high-cadence but this system.} and the additional target, V1208\,Sco. The spectra of HD\,154407C (see below) were used to verify that the RV variations were coherent with the photometric variations, and because they were, the target was kept. To ensure that the analyzed curves suffer as little as possible from crowding, we relied on the PDCSAP curves because they are supposed to be corrected for these effects. Three targets lack 2\,min data, however: HD\,350685, LS\,I\,+61\,275, and V1208\,Sco. For these targets, we extracted the \te\ light curves with the \te-GAIA light curve (TGLC) package\footnote{https://github.com/TeHanHunter/TESS\_Gaia\_Light\_Curve/blob/ main/tutorial/TGLC\_tutorial.ipynb} \citep{tglc}. This package uses neighboring sources in the {\it Gaia} catalog, assuming they are constant. We choose the aperture extraction\footnote{PSF extractions were more noisy, as usual, and calibrated curves were detrended for daily changes, which we investigated, and these data were therefore not used here.} to derive the cleaned photometry. 

{\it Kepler} data are also available from the MAST portal for TYC\,1881-933-1 (see `K' in the last column of Table \ref{targetlist}). As for \te\ data, we kept only the high-quality data points of the PDCSAP curve and then converted its fluxes into magnitudes. A long-term trend was fit by a straight line and then taken out, so that the {\it Kepler} and \te\ curves display the same base level. 

\subsection{Spectroscopy}
While eclipses with reflection provide much information on the systems, complementary spectroscopy is required to assess the true nature of the systems (pre/post-interaction). We therefore gathered spectroscopic data for all the systems.

For three systems, the velocities are available in the literature. LS\,I\,+61\,275 was studied by the Georgia state team \citep{hua06b,hua06} and by the Apache Point Observatory Galactic Evolution Experiment (APOGEE, \citealt{jon20}). The former references provided effective temperature, projected rotational velocity, and two radial velocity (RV) measurements. The latter reference provides 8 RV points. While both indicate that the RVs decrease around the main eclipse, they suggest very different RV amplitudes: about 100\,\kms\ for the former ones, and four times lower for the latter (Fig. \ref{rvphot}). This strong discrepancy on the RV amplitudes, coupled with the impossibility to derive a good photometric fit when using a mass function derived from the more numerous APOGEE data, prevented us from performing a detailed analysis of this star before a new high-quality spectroscopic campaign is performed. TYC\,1881-933-1 was studied in the framework of the Large Sky Area Multi-Object Fibre Spectroscopic Telescope (LAMOST) survey. Corrected RVs, determined from the blue and red channels, were published by \citet{zha22}. The blue-channel RVs behave similarly to the red channel RVs, but have more scatter. We therefore only kept the latter to calculate an orbital solution. One value, at BJD\,$=2\,458\,122.20438$, appears to be discrepant and was therefore taken out, which left 43 RV points. Effective temperature, gravity, and projected rotational velocity were estimated for each LAMOST spectrum by \citet{xia22}. The average values and their dispersions are reported in Table \ref{targetlist}. Finally, the velocities for the primary of V1208\,Sco were provided by \citet{ban22}, and its effective temperature, gravity, and projected rotational velocity were evaluated by \citet{mcs09}. These literature velocities are recalled in the appendix (Table \ref{litrv}). 

We collected spectroscopy for the other targets using several dedicated campaigns. Four spectra were collected for HD\,350685 with the Hermes echelle spectrograph installed on the Mercator telescope (La Palma Observatory). In the framework of the Optical Infrared Coordination Network (OPTICON) program, spectra were collected for us with the SOPHIE\footnote{Spectrographe pour l'Observation des PH\'enom\`enes des Int\'erieurs stellaires et des Exoplan\`etes} echelle spectrograph at the Haute-Provence Observatory for HD\,254346 (four spectra; program 23B004, PI Naz\'e) and with the CARMENES\footnote{Calar Alto high-Resolution search for M dwarfs with Exoearths with Near-infrared and optical Echelle Spectrographs} echelle spectrograph at the Calar-Alto Observatory for LS\,I\,+61\,145 and TYC\,741-1565-1 (two spectra each; program 24B010, PI Naz\'e). Finally, between five to seven spectra were collected at the European Southern Observatory (ESO) with the XShooter spectrograph for AN\,Dor, TYC\,741-1565-1, HD\,112485, V1061\,Cen, and HD\,154407C under run ID=114.26ZK.001 (PI Naz\'e). These spectra were complemented by archival spectra available in the ESO archives\footnote{https://archive.eso.org} for HD\,309317 (eight FEROS\footnote{Fiber-fed Extended Range Optical Spectrograph} spectra) and AN\,Dor (six XShooter and three FEROS spectra). The XShooter spectra were taken with a medium spectral resolution ($R$ between 5500 and 10\,000), and all other data relied on high-resolution spectrographs ($R$ between 40\,000 and 95\,000). All spectra were provided in reduced form by the observatories. The CARMENES spectra were calibrated for vacuum wavelengths, however, while the other data used an air wavelength calibration. The CARMENES wavelengths were therefore converted using $\lambda_{air}=\lambda_{vac}/1.000272$. Cosmic rays were eliminated using a median filter of size 2 ({\sc MIDAS filter/median} command) for all but XShooter and CARMENES spectra (which had very short exposures and hence no cosmic-ray hits).

\section{Results}
\subsection{Light-curve analyses}
First, we performed a period search on all individual light curves up to 25\,d$^{-1}$ to assess their frequency content \citep{hmm}. The orbital signal clearly dominated in all cases. All \te\ curves of the same target were then merged, and the period search was repeated around the main signal, with a smaller frequency step (adapted to the longer time interval covered by the data). This allowed us to obtain preliminary values for the orbital period and reference time of the main deeper eclipse, which were then refined by examining the folded light curves. The final best-fit ephemerides can be found in Table \ref{phot1}. An average phase-folded light curve (with 200 phase bins for \te\ data and 100 for {\it Kepler} data) was then calculated from all crowding-corrected curves, except when there was too much noise or some obvious change between individual light curves (the data we used are identified by boldface in Table \ref{targetlist}). These average curves were used to remove the binary signal from the data, and the residuals were examined in search of pulsations. The appendix provides the results of this search.

The average binned light curves were fit using the model by \citet{moe15}. This analytical model uses two Gaussians to represent the eclipses and a cosine function to reproduce the reflection effect: $\langle I \rangle + \Delta I_1 [\exp^{-\phi^2/2 \theta_1^2)}+\exp^{(\phi-1)^2/2 \theta_1^2}] + \Delta I_2 \exp^{-(\phi-\phi_2)^2/2 \theta_2^2} + 0.5 \Delta I_{refl} * [\cos (2 \pi (\phi-1/2)) +1]$, where the numbers $i=1$ and 2 correspond to the deeper and secondary eclipses, respectively, $\Delta I_i$ and $\theta_i$ are the eclipse depths and widths, respectively, $\Delta I_{refl}$ is the amplitude of the reflection, $\phi_2$ is the phase of the second eclipse, and $\phi$ is the orbital phase derived from the average ephemeris (see Cols. 2 and 3 of Table \ref{phot1}). The best-fit parameters, found through a Levenberg-Marquardt algorithm, are listed for each target in Table \ref{phot1}. Despite its simplicity, this model is very useful to catch the main features of the light curves: the eclipse depth, the reflection strength, and the potential eccentricity. For example, the phase $\phi_2$ of the shallower eclipse is compatible with 0.5 for all but three cases: AN\,Dor, HD\,254346, and V1061\,Cen. This indicates that these three systems have eccentric orbits. Their lowest eccentricities, $e_{min}=\pi |\phi_2-0.5|/2$ \citep{moe15}, are 0.049, 0.010, and 0.008, respectively (and the typical values should be about 1.6 times $e_{min}$ for a uniform distribution of the periastron angle $\omega$).

The model of \citet{moe15} was then used to precisely derive the individual eclipse times. For each orbital cycle covered by the data, the times of the main and secondary eclipses were predicted from the average ephemerides and the phase of secondary eclipse (Table \ref{phot1}). Then, the eclipse models were shifted within the range $\pm2\times \theta_i$ (with $i=1,2$ for the main and secondary eclipses, respectively), and the $\chi^2$ between light curve and model was evaluated for each shift within a phase interval of $\pm3\times \theta_i$ around each eclipse. We then examined not only the differences between the best-fit eclipse times and their predictions but also the interval between primary and secondary eclipses of the same cycle. This allowed us to detect apsidal motion in AN\,Dor and global eclipse shifts in  TYC\,1881-933-1, HD\,309317, and V1061\,Cen. Dedicated analyses of these stars are provided in Sections 3.4 and 3.5, respectively.

\subsection{Spectral analyses}
Two independent derivations of the RVs were made for the spectra. In the first derivation, normalization was achieved by spline fitting through dedicated spectral windows, and then RVs of H$\beta$ and He\,{\sc i}\,$\lambda$5876\AA\ were evaluated independently by cross-correlation with a 20kK model spectrum from the BSTAR2006 database \citep{lan07}. Finally, an average of the two RVs was calculated. In the second derivation, normalizations were made for the whole wavelength range using the tool called spectrum normalization neural network (SUPPnet) by \citet{roz22}\footnote{https://rozanskit.com/suppnet/}. The RVs were then derived using the iSpec tool \citep{bla19}\footnote{https://github.com/marblestation/iSpec}. To do this, iSpec cross-correlated the observed spectra with a generic TLUSTY model for an early-B star ($T_{eff}=20$\,kK, $\log(g)=4$, and $v\sin(i)=150$\,\kms), taken from the BSTAR2006 database\footnote{downloaded from https://lweb.cfa.harvard.edu/$\sim$sblancoc/iSpec/grid/}. The wavelength ranges were 3650--5600\AA\ (XShooter UV subspectra), 3800--6800\AA\ (FEROS and Hermes spectra), 3900--6800\AA\ (SOPHIE spectra), and 5700--6800\AA\ (CARMENES spectra). The derived RVs, provided in Table \ref{rv}, agreed well with those derived from the previous method. Finally, we also made a trial with templates of lower temperatures (8\,kK and 14\,kK, which bracket the derived secondary temperatures; see below), but to no avail. The results for the 14\,kK template were similar to those of the 20\,kK template, and the 8\,kK template resulted in velocities with extremely large errors and/or that were constant (or affected by the primary). 

The individual spectra of each target were shifted by minus the determined RVs to place them in the same zero velocity reference frame, and they were then averaged. A fitting of these average spectra was then performed within iSpec to determine the closest BSTAR2006 TLUSTY models and thereby obtain the best physical parameters of the (primary) star. This was done first with the metallicity set free to vary and over the whole wavelength range. The fitting was then repeated with the metallicity fixed to solar and considering only the ranges 4000--5100\AA, 5860--5885\AA, and 6540--6700\AA, in which the most interesting and most numerous spectral lines appear. The best-fit values for the effective temperature $T_{eff}$, gravity $\log(g)$, and projected rotational velocity $v\sin(i)$ are provided in Table \ref{targetlist}, except for LS\,I\,+61\,145, for which the CARMENES red spectra contained too few stellar lines for a meaningful estimate of the effective temperature and gravity. The best-fit temperatures were used, as mentioned above, to improve the evaluations of the spectral types and bolometric corrections. Finally, the bolometric luminosities, effective temperatures, gravities, and projected rotational velocities were entered in BONN Stellar Astrophysics Interface (BonnSAI)\footnote{https://www.astro.uni-bonn.de/stars/bonnsai/index.php} \citep{sch14} to derive estimates of the evolutionary masses, radii, and ages of the massive primaries (Table \ref{targetlist}). 

\begin{figure}
  \begin{center}
    \includegraphics[width=8cm]{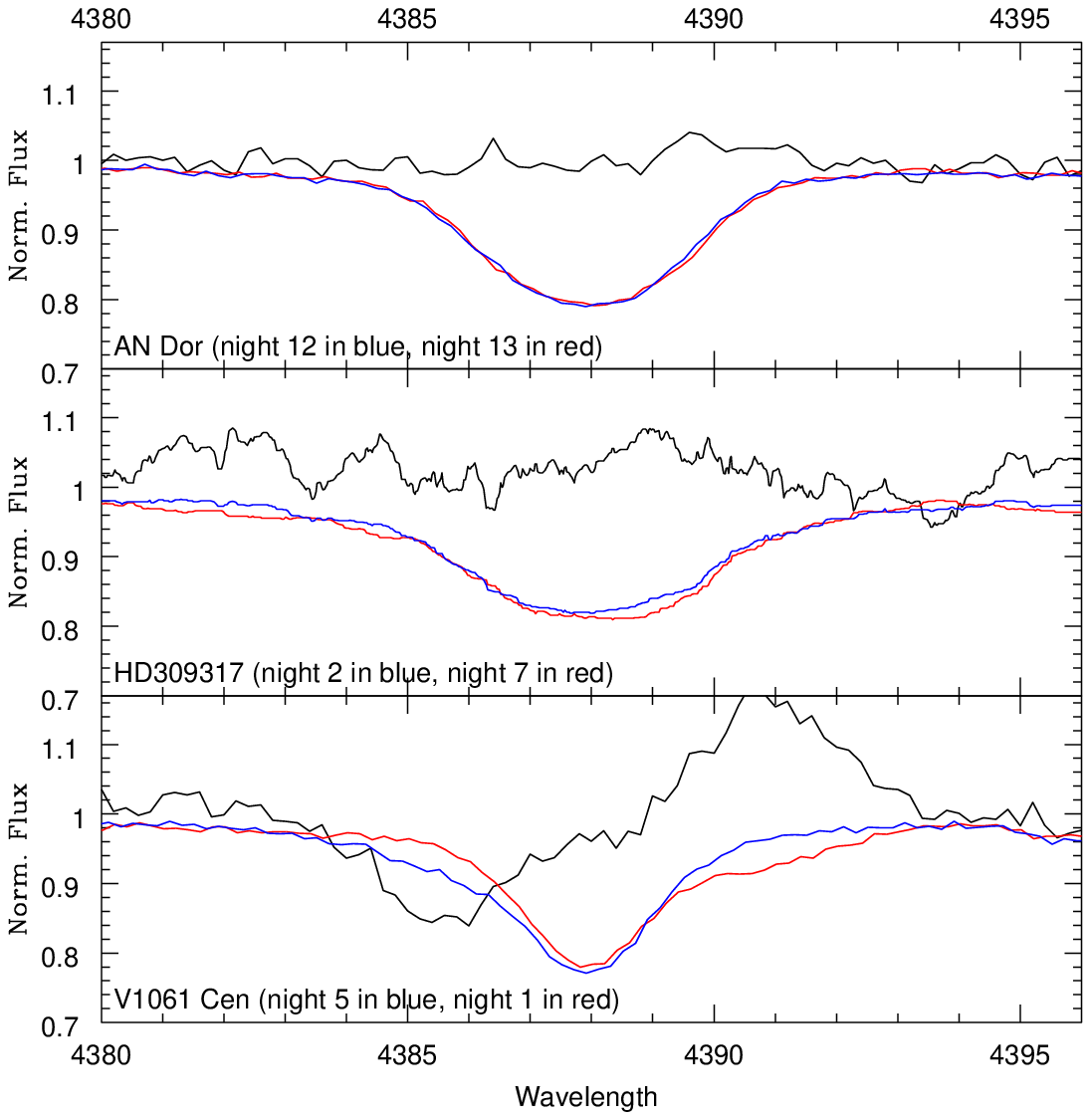}
  \end{center}  
  \caption{Line profiles of He\,{\sc i}\,$\lambda$4388\AA\ in the most blue- and redshifted spectra (blue and red lines, respectively) for AN\,Dor, HD\,309317, and V1061\,Cen. The spectra were corrected for the measured RVs, and the black line provides their difference (multiplied by four and then shifted by +1 to facilitate the comparison with the lines). \label{lineprof}}
\end{figure}

At first sight, the spectra only showed one stellar signature. To assess whether a weak second component existed, we shifted the spectra with minimum and maximum RVs to the zero-velocity reference frame. We plotted the superimposed spectral lines and also calculated their difference. Except for the case of V1061\,Cen (which is further analyzed in Sect. 3.6), the line profile appeared to be constant, without any weak additional component. The targets thus appear to be of SB1 type in our dataset. Most probably, more spectra of the highest quality are needed to detect the secondary signature, as for HD\,25631 \citep{naz23}. \citet{tka24} concluded from the same FEROS dataset as ours, however, that HD\,309317 was SB2. Figure \ref{lineprof} compares the line profiles of AN\,Dor, HD\,309317, and V1061\,Cen after the spectra were shifted in a zero-velocity reference frame. Only the latter target clearly shows the signature of a second component in the wings of the main line. Other lines and other phases were also checked, but we were still unable to detect any signature typical of a SB2 nature for HD\,309317 (e.g., a second component or line doubling, a shallower and broader line at quadratures or a deeper and narrower line at conjunctions). In this context, we also note that the $K_1$ and $K_2$ mentioned by \citet{tka24} would lead to a mass ratio of 0.44, implying a late-B star nature for the secondary, whose lines should then be rather obvious in the data (but even a disentangling trial failed to unveil a meaningful secondary spectrum). Therefore, within the limitation of current data, the SB2 classification must be revised to SB1. 

\subsection{System characterization}

After applying the methods outlined in Sections 3.1 and 3.2, we present the folded light curves and RVs in Fig. \ref{rvphot} for all targets. One common characteristic then appears obvious. Around the time of the deepest eclipse, the RV of the massive star is found to decrease. This implies that the deepest eclipse corresponds to the eclipse of the massive star by its companion (i.e., a conjunction with the massive star behind its companion), and this massive star is the hotter component of the system. Our targets thus possess the same nature. They are young pre-interacting systems with fast-rotating B-type primaries and very low-mass secondaries.

With light curves and RVs at hand, the physical parameters of the systems were constrained in detail. For V1208\,Sco, a SB2 fitting was performed by \citet{ban23}. The velocity amplitudes amount to $K_1=44.7\pm0.4$\,\kms\ and $K_2=259.6\pm0.8$\,\kms\ (L. Mahy, private communication), which leads to a mass ratio of $q=0.172\pm0.002$ and to mass estimates of $M_1\sin^3(i)=12.99\pm0.11$\,M$_{\odot}$ and $M_2\sin^3(i)=2.24\pm0.03$\,M$_{\odot}$. For the other targets, we started by fitting the RVs with sinusoids, except for AN\,Dor (see the subsection below). Only three systems were found to be eccentric, and two of them (HD\,254346 and V1061\,Cen) have a low $e_{min}$, together with only a few spectra. In these conditions, the eccentricity cannot be constrained precisely enough, and a noneccentric solution provides a sufficiently good fit for our purposes. Table \ref{binsol} yields the best-fit parameters, and we note that our RV amplitude of HD\,350685 agrees well with that published by \citet{tka24}. Finally, the RV solution for LS\,I\,+61\,145 should be considered preliminary because only two RV points are available.

From the RV amplitude and the period, we calculated the mass function . Using the primary masses from Table \ref{targetlist}, we derived first estimates of the secondary mass and the mass ratio $q$ for an inclination $i=70-90^{\circ}$. The average light curves were then fit using Nightfall v2.0.2 \citep{wic11}, assuming $I$ band for the data (the closest to \te\ bandpass), detailed reflection, and a quadratic limb darkening. The eccentricity, mass ratio, plus the mass and effective temperature of the primary star were fixed to the values appropriate for the target under consideration. A preliminary fitting provided a first inclination value, which was iteratively used, in conjunction with the mass function, to refine the mass ratio. The final best-fit parameters are provided in Table \ref{binsol}. One caveat must be mentioned. The errors in this table correspond to formal errors around the provided solution. Exploration of the parameter space reveals that solutions of similar quality but with different parameters can be found. As in \citet{naz23}, the true uncertainty is thus larger than that around a specific individual solution. The realistic error values are a few degrees for inclinations, about 10\% on radii, and several kK for the temperature of the secondaries.

We used the effective temperatures and radii to evaluate the bolometric luminosities. In all cases, the secondary appears to be much fainter because their luminosities are $<2$\% of the primary luminosities. Furthermore, the derived total luminosities agree well with those derived from the $V$ magnitude, distance, and extinction (Table \ref{targetlist}). The radii of the primaries also agree with those found from evolutionary tracks (Table \ref{targetlist}). Finally, the gravities of the primaries were computed from the (assumed) mass and the derived radius. A good agreement was again found with those estimated from the spectral fitting by atmosphere models (Table \ref{targetlist}). With values around $\log(g)=4$, the gravities also confirm the main-sequence status of the primaries.

\begin{figure}
  \begin{center}
    \includegraphics[width=8cm]{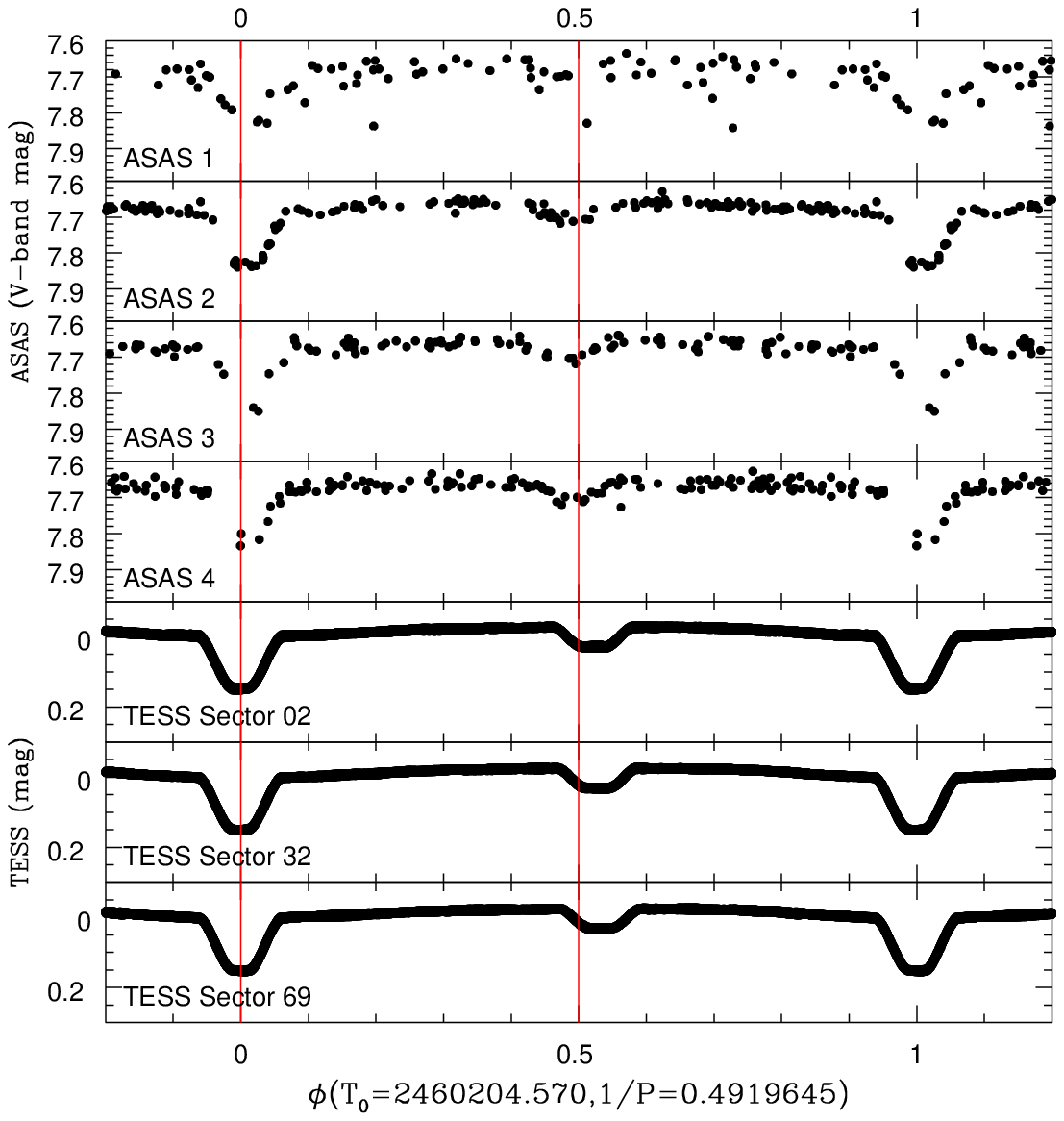}
  \end{center}  
  \caption{Light curve observed with {\it ASAS} (four epochs; see text) and \te\ for AN\,Dor (three sectors shown). The separation between the primary and secondary eclipses evolves. \label{andorLC}}
\end{figure}

\subsection{Apsidal motion in AN\,Dor}
The intervals between the main and secondary eclipses remain compatible with a constant for all targets but one: AN\,Dor. Increasing shifts between predicted and observed eclipse times were also observed for this system. All this is typical of apsidal motion \citep{Gim95,Cla21}. To obtain a longer temporal baseline, that is, to improve the characterization of this effect, we also downloaded All-Sky Automated Survey ({\it ASAS}\footnote{https://www.astrouw.edu.pl/asas/?page=aasc}, \citealt{poj97}) data of this system and kept only data points with the highest quality (flag=`A'). We then cut the {\it ASAS} curve into four parts (HJD$-2\,450\,000 <2350$, 2350--3200, 3200--4000, and $>4000$). For each interval, we first estimated the shift between the folded {\it ASAS} data and the average binned \te\ curve within $\Delta(\phi)=\pm0.15$ of the main eclipse. After we determined the best-fit time of the main eclipse, the Moe \& Di Stefano model was fit to the folded {\it ASAS} data to derive the phase (hence, the time) of the secondary eclipse. The evolving light curve of AN\,Dor is shown in Fig. \ref{andorLC}, and the best-fit model parameters are provided at the bottom of Table \ref{phot1}, except for the first interval, which lacks enough data points to provide meaningful results. 

Newtonian apsidal motion is a secular change in the longitude of the periastron $\omega$ that is caused by the nonspherical shape of the gravitational potential in an eccentric close binary \citep[e.g.,][and references therein]{Gim95,Sch16,Ros20a}. \citet{Gim95} provided the mathematical expressions of the times of photometric minima as a function of $e^n \cos{n\,\omega}$ and $e^n \sin{n\,\omega}$, where $e$ is the eccentricity, and $n$ takes integer values between 1 and 6. These quantities are multiplied by functions of $\cot{i}$ and $\csc{i}$, where $i$ is the orbital inclination.

\begin{figure}
  \begin{center}
    \includegraphics[width=8cm]{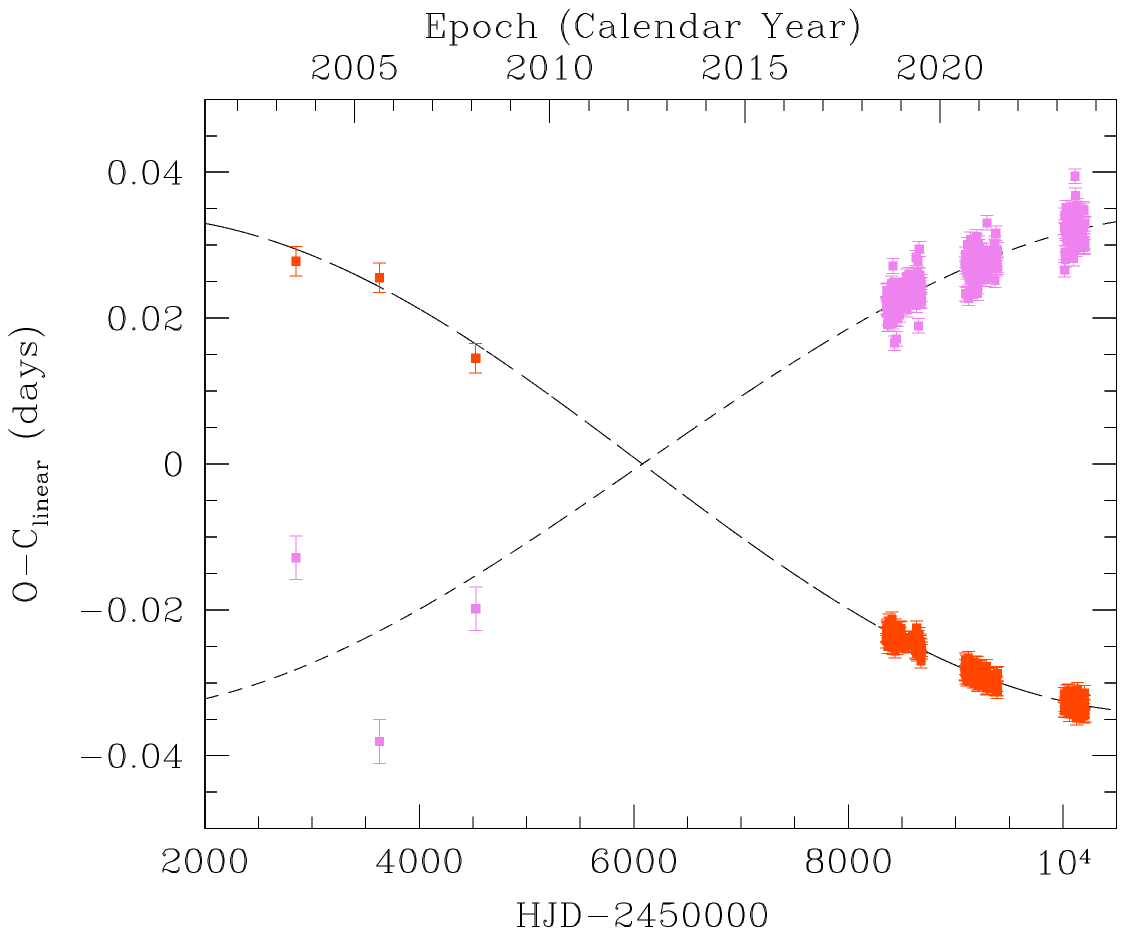}
  \end{center}  
  \caption{Best-fit adjustment to the differences between the observed times of minima of AN\,Dor and the linear ephemerides $O-C_{\rm linear}$ for $i = 80^{\circ}$. The orange and pink symbols show the eclipses by the primary and secondary stars, respectively. Data prior to 2010 are taken from the {\it ASAS} survey, and data after 2018 are taken from \te\ photometry. The long and short dashed lines yield our best-fit theoretical relations for the times of minima. In this plot, the linear ephemerides are $T_0=2\,460\,204.6031$ and $P^{sid}=2.0326782$\,d.  \label{OC}}
\end{figure}

We used 469 times of minima inferred from the \te\ data of AN\,Dor along with 6 times of minima derived from the {\it ASAS} observations to search for apsidal motion. For this purpose, we adopted relation (15) of \citet{Gim95} to compute the theoretical times of minima and compared them to the observed values of the difference with respect to best-fit linear ephemerides ($O-C_{\rm linear}$). In addition to the orbital inclination $i$, the theoretical expression of $O-C$ depends on four parameters: the anomalistic orbital period $P^{an}$, the eccentricity $e$, the rate of apsidal motion $\dot{\omega}$, and the longitude of periastron $\omega_0$ at a reference time. For values of $i$ between $70^{\circ}$ and $90^{\circ}$, by steps of $5^{\circ}$, we built a four-dimensional grid to sample the parameter space. For each combination of parameters of our grid, we adjusted the linear ephemerides in such a way as to achieve a zero mean residual. When we identified a combination of parameters that yielded the lowest $\chi^2$ residuals between the theoretical and observed $O-C_{\rm linear}$ values, we refined the grid to achieve a higher resolution. As a general result, we found that the dependence of the solutions on $i$ is very weak in the range of inclinations we considered. Only the best-fit $e$ and $\omega_0$ parameters changed slightly as a function of $i$. The largest differences were found in the best-fit eccentricity, which increased from $0.0496$ to $0.0528$ as $i$ changed from $70^{\circ}$ to $90^{\circ}$. Table\,\ref{omegadot_photo} lists the best-fit parameters and their 1$\sigma$ error bars for $i = 80^{\circ}$, and the corresponding best-fit adjustment of time series of $O-C_{\rm linear}$ values is displayed in Fig.\,\ref{OC}. 

\begin{table}
  \caption{Best-fit parameters inferred from the times of photometric minima of AN\,Dor assuming $i = 80^{\circ}$.\label{omegadot_photo}}
  \begin{center}
  \begin{tabular}{l c}
    \hline
    Parameter & Value \\
    \hline
    \vspace*{-2mm}\\
    $P^{an}$ (d) & $2.0328825^{+0.0000125}_{-0.0000165}$ \\
    \vspace*{-2mm}\\
    $e$ & $0.0520^{+0.0020}_{-0.0015}$ \\
    \vspace*{-2mm}\\
    $\dot{\omega}$ ($^{\circ}$\,yr$^{-1}$) & $6.50^{+0.45}_{-0.50}$ \\
    \vspace*{-2mm}\\
    $\omega_0$ ($^{\circ}$) & $343.4^{+3.6}_{-4.7}$ \\
    \vspace*{-2mm}\\
    $\gamma$ (km\,s$^{-1}$) & $54.9\pm2.0$\\
    \vspace*{-2mm}\\
    $K_1$ (km\,s$^{-1}$) & $59.6^{+2.8}_{-2.7}$ \\
    \vspace*{-2mm}\\
    $T_0^p$ (HJD-2460000) & $204.0095^{+0.0105}_{-0.0105}$ \\
    \vspace*{-2mm}\\
    \hline
  \end{tabular}
  \end{center}
  \tablefoot{$P^{an}$ stands for the anomalistic orbital period, $T_0^p$ is the time of periastron passage, and $\omega_0$ yields the value of $\omega$ at the time $T_0^p$. }
\end{table}

Given the high number of \te\ data points and their excellent accuracy, our best fit is clearly dominated by these points. The inclusion of the {\it ASAS} data still provided a valuable confirmation of the inferred parameters. The times of primary eclipses derived from the {\it ASAS} data agree very well with our best solution. The secondary eclipses are subject to much higher uncertainties in the {\it ASAS} observations, but even there, the trend is consistent with our solution. From what we find here, the rate of apsidal motion corresponds to a period $U = (55.4 \pm 4.3)$\,yr.

\begin{figure}
  \begin{center}
    \includegraphics[width=8cm]{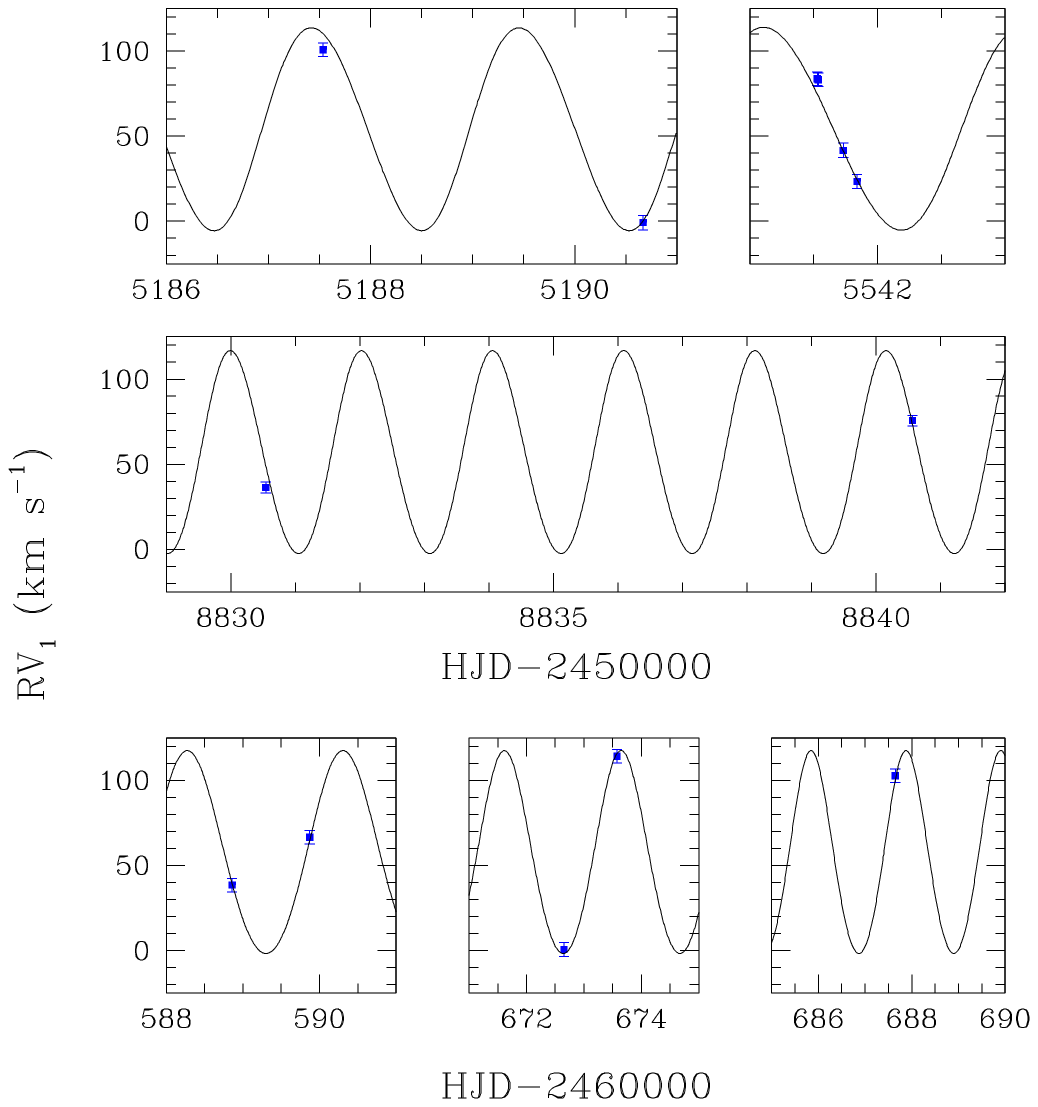}
  \end{center}  
  \caption{Comparison between the measured RVs of AN\,Dor (blue symbols with error bars) and the best-fit orbital solution explicitly accounting for apsidal motion. \label{ANDorRVfit}}
\end{figure}

Keeping the eccentricity, rate of apsidal motion and anomalistic orbital period fixed to their best-fit values derived from our analysis of the photometric data, we then adjusted a RV curve that explicitly accounted for apsidal motion \citep[see Eqs.\,(1) and (2) of][]{rau16b} to determine the time of periastron passage and the amplitude of the SB1 RV curve. The results are given in Table\,\ref{omegadot_photo} and illustrated in Fig.\,\ref{ANDorRVfit}.

Because of the apsidal motion, we did not combine the whole set of \te\ light curves. We instead averaged only a subset of curves from Sectors 63, 64, and 65, which cover HJD$=2\,460\,014-97$. At the time (Table\,\ref{omegadot_photo}), the periastron argument should be around $\omega=340.76^{\circ}$. We therefore calculated a photometric solution as for other stars, but fixing the periastron argument to this value, $e=0.052$ (Table\,\ref{omegadot_photo}), $T_{eff}^1=21529$\,K, $M_1=8.2$\,M$_{\odot}$ (Table \ref{targetlist}), and $q=0.205$ (found from the mass function and the inclination of a preliminary fitting run). The best-fit parameters are provided in Table \ref{binsol}.

\begin{figure*}
  \begin{center}
    \includegraphics[width=9cm]{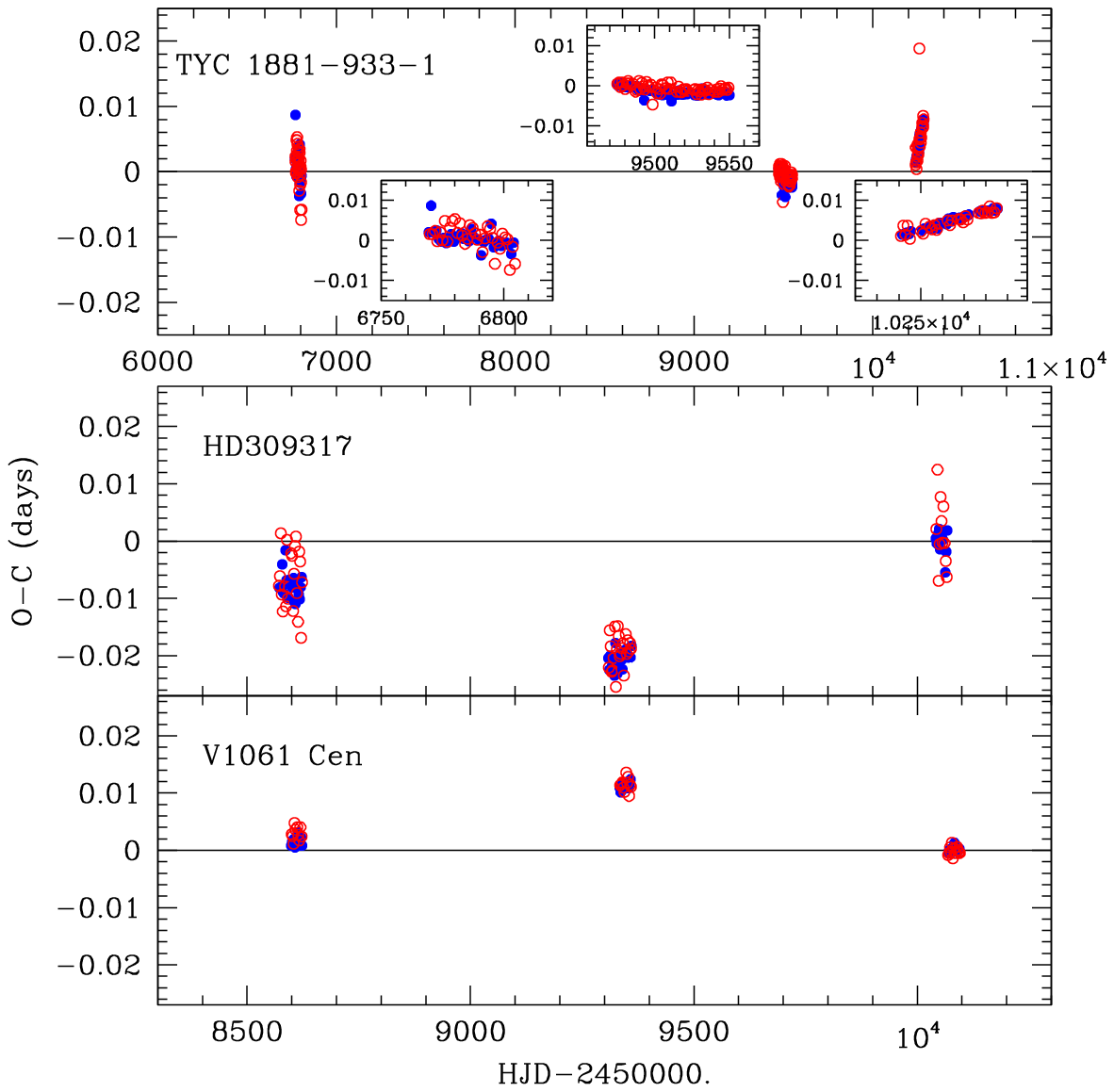}
    \includegraphics[width=9cm]{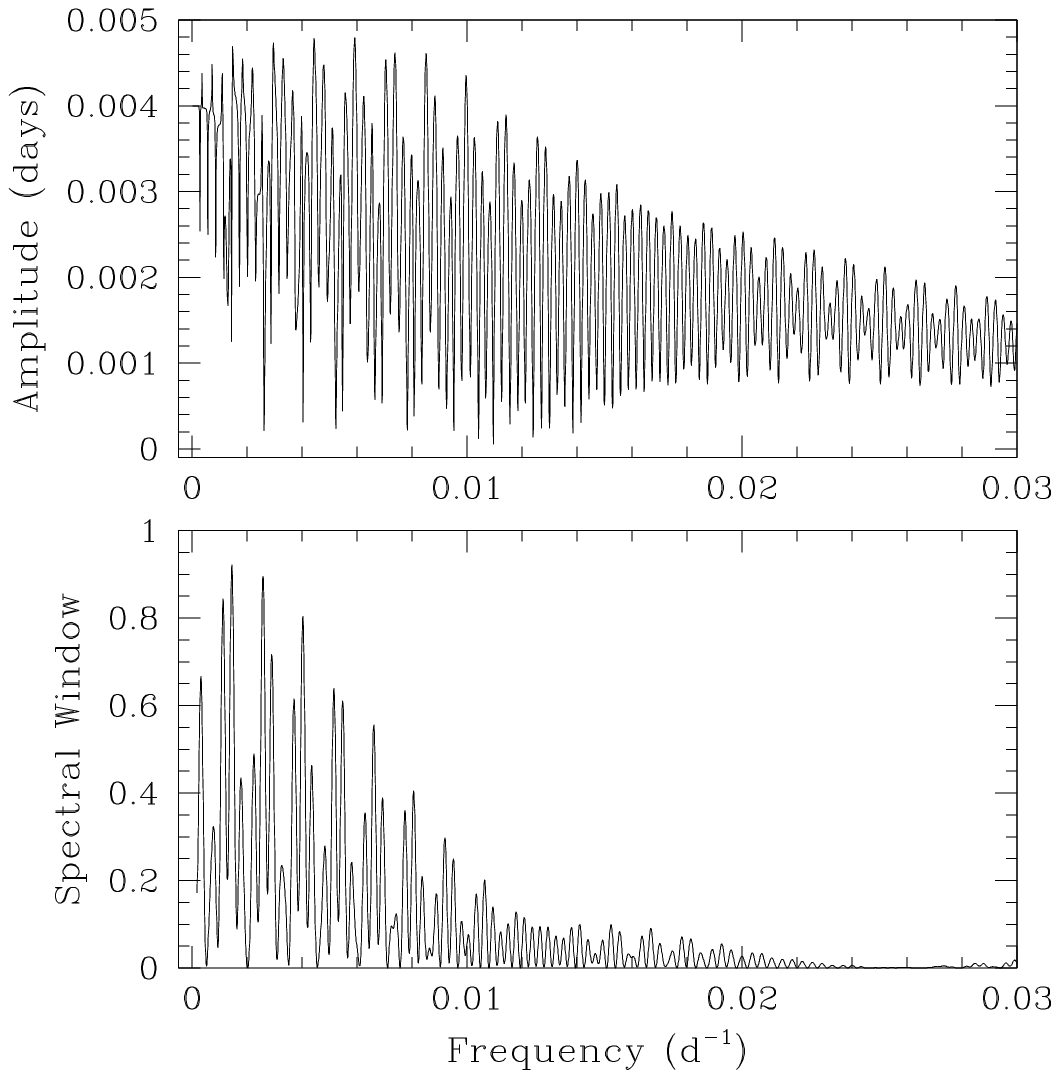}
  \end{center}  
\caption{{\it Left:} Differences between observed and predicted eclipse times based on the ephemerides of Table \ref{phot1}. The data for the deeper eclipse are shown by blue dots, and the data for the shallower eclipse are shown as red circles. Close-ups are provided for TYC\,1881-933-1. {\it Right:} Fourier spectrum of the $O-C$ values of the times of primary minimum of TYC\,1881-933-1 (top panel) and the associated spectral window (bottom panel). \label{SpOC}}
\end{figure*}

\subsection{Eclipse time shifts}
Even when the interval between the two eclipses remained stable, evolving shifts of the light curves can be detected. This was so for three cases: TYC\,1881-933-1, HD\,309317, and V1061\,Cen (Fig. \ref{SpOC}). The most likely explanation for this change is a light-travel time effect (LTTE). In a triple system, the motion of the inner eclipsing binary about the center of mass of the triple system leads to a periodic change in the eclipse times because the velocity of light is finite. This is expressed as
\begin{equation}
  (O-C)(t) = \frac{a_{bin}\,\sin{i}}{c}\,\frac{1 - e^2}{1 + e\,\cos{\psi(t)}}\,\sin{(\psi(t) + \omega)}
  \label{ltte} ,
\end{equation}
where $a_{bin}$, $e$, $i$, and $\omega$ stand for the semimajor axis, eccentricity, inclination, and argument of periastron of the orbit of the inner binary around the center of mass of the triple system. For each epoch $t$, the true anomaly $\psi(t)$ of the inner binary was computed solving Kepler's equation, which involves $e$, the time of periastron passage $T_0^p$, and the orbital period $P_{\rm out}$ of the outer orbit. The comparison of the observed $O-C$ values with Eq.\,(\ref{ltte}) then allowed us to constrain several parameters of the outer orbit. These are $\frac{a_{bin}\,\sin{i}}{c}$, $e$, $\omega$, $T_0^p$, and $P_{\rm out}$.

For each of the three targets, three epochs of photometric observations exist. For HD\,309317 and V1061\,Cen, the data span a total duration of 1500\,d, and the differences are mostly detected between the second epoch and the other two (which have similar $O-C$). The $O-C$ changes in that second epoch by about 0.02\,d and 0.01\,d, respectively. With basically only three points available, the outer orbit is clearly not well constrained. Because the inner binary details remain unknown for V1061\,Cen (see Section 3.6), we only provide a preliminary estimate for HD\,309317. For the latter system, the orbital solution suggests the binary mass to be about 10\,M$_{\odot}$ (Table \ref{binsol}). We further assumed a zero eccentricity for the outer orbit and inclinations between 30$^{\circ}$ and 90$^{\circ}$. We also considered two (simple) possibilities: either the observations show half a cycle or a full cycle, that is, $\frac{a_{bin}\,\sin{i}}{c}=0.02$\,d and $P\sim3000$\,d or $\frac{a_{bin}\,\sin{i}}{c}=0.01$\,d and $P\sim1500$\,d. Kepler's law indicates $(a_{bin}\sin(i))^3=\left(\frac{P}{2\pi}\right)^2\frac{GM_3^3\sin(i)^3}{(M_{bin}+M_3)^2}$. Therefore, $M_3$ is found to be 4--14\,M$_{\odot}$. While RVs measured on spectra taken over only 60\,d should not be affected by the outer orbit, the spectral lines from a star at the higher end of this mass interval should have been detected in the spectra of HD\,309317, but they only show a single component (Fig. \ref{lineprof}). The coverage of the outer orbit is scarce, however. More observations and therefore a more precise derivation of the outer orbit are now needed to reach a better knowledge of the tertiary component. 

\begin{table*}
  \caption{Best-fit parameters of the outer orbit inferred from the times of primary minima of TYC\,1881-933-1.\label{solltte}}
  \begin{center}
  \begin{tabular}{l c c c c}
    \hline
    Parameter & $P_{\rm out} = 675.2^{+3.6}_{-3.7}$\,d &  $P_{\rm out} = 337.8^{+1.5}_{-1.2}$\,d &  $P_{\rm out} = 225.2^{+0.9}_{-0.7}$\,d & $P_{\rm out} = 169.2^{+0.4}_{-0.6}$\,d \\
    \hline
    \vspace*{-2mm}\\
    $\frac{a_{bin}\,\sin{i}}{c}$ ($10^{-3}$\,d) & $13.9^{+4.1}_{-3.1}$ & $8.8^{+2.2}_{-2.3}$ &  $6.5^{+1.6}_{-1.5}$ & $5.4^{+1.3}_{-1.2}$ \\
    \vspace*{-2mm}\\
    $e$ & $0.71^{+0.05}_{-0.05}$ & $0.53^{+0.12}_{-0.08}$ & $0.38^{+0.17}_{-0.13}$ & $0.21^{+0.22}_{-0.11}$ \\
    \vspace*{-2mm}\\
    $\omega$ ($^{\circ}$) & $249^{+26}_{-14}$ & $255^{+30}_{-30}$ & $236^{+64}_{-26}$ & $240^{+90}_{-40}$  \\
    \vspace*{-2mm}\\
    $T_0^p$-2450000 & $9503^{+22}_{-3}$ & $9512^{+13}_{-17}$ & $9501^{+34}_{-11}$ & $9508^{+42}_{-18}$ \\
    \vspace*{-2mm}\\
    rms ($10^{-3}$\,d) & $7.4$ & $7.4$ & $7.4$ & $7.4$ \\
    \hline
  \end{tabular}
  \end{center}
\end{table*}

\begin{figure*}
  \begin{center}
    \includegraphics[width=4.5cm]{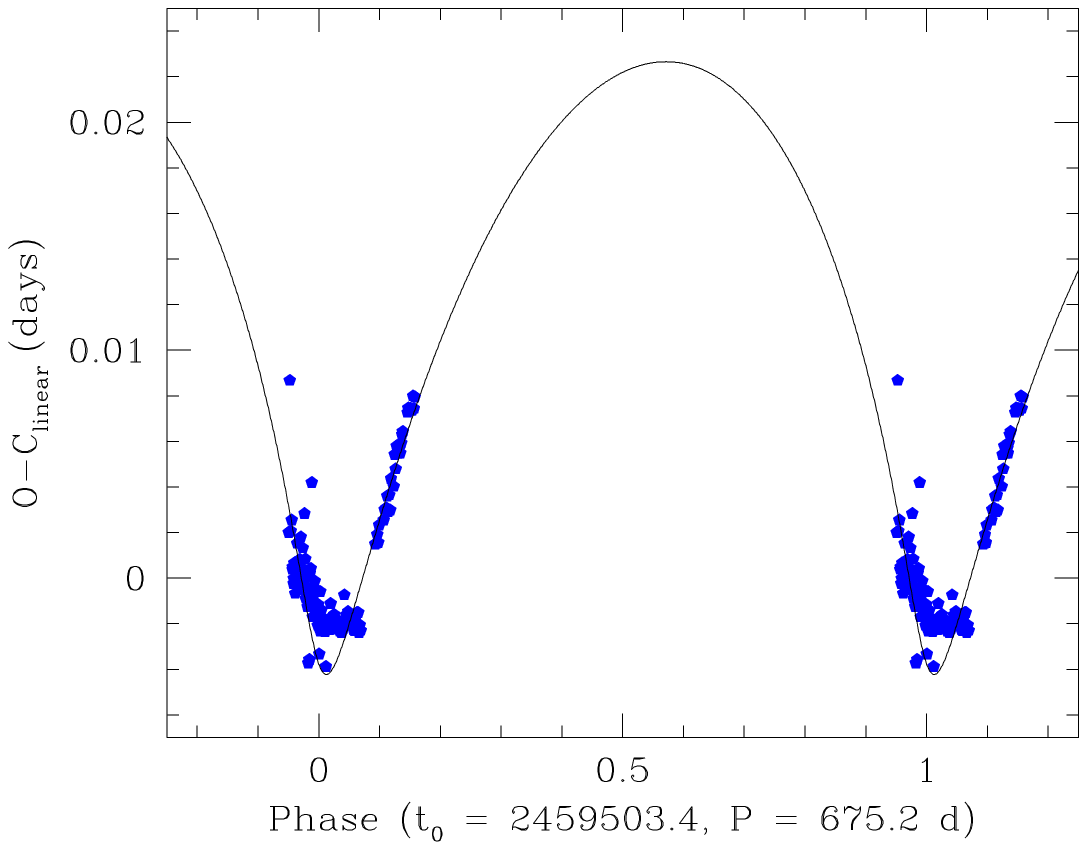}
    \includegraphics[width=4.5cm]{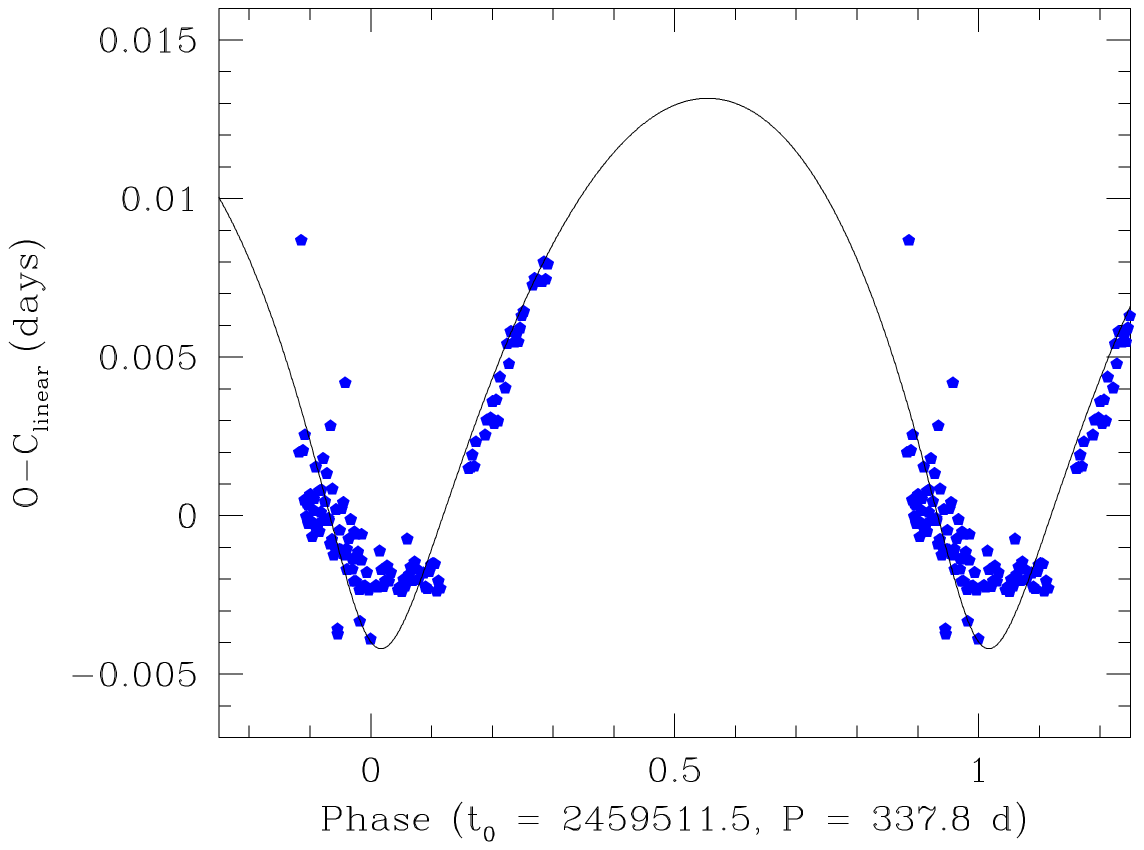}
    \includegraphics[width=4.5cm]{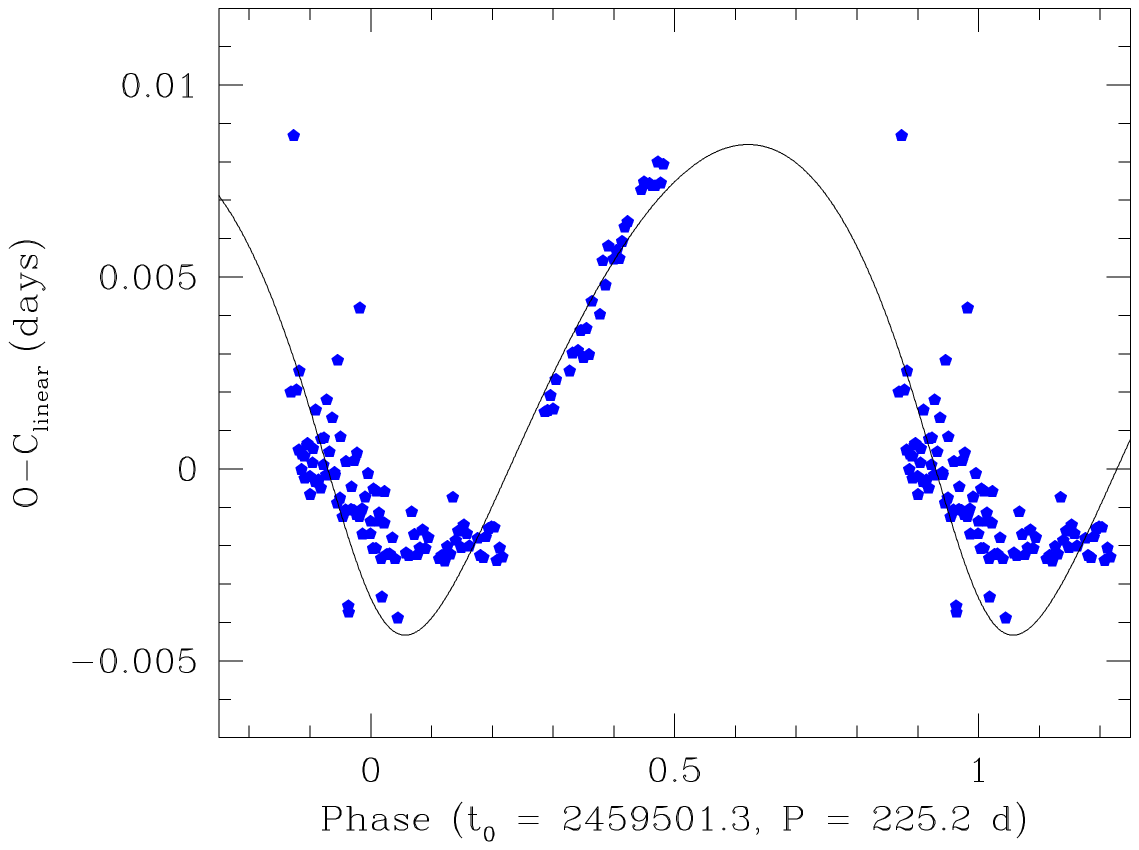}
    \includegraphics[width=4.5cm]{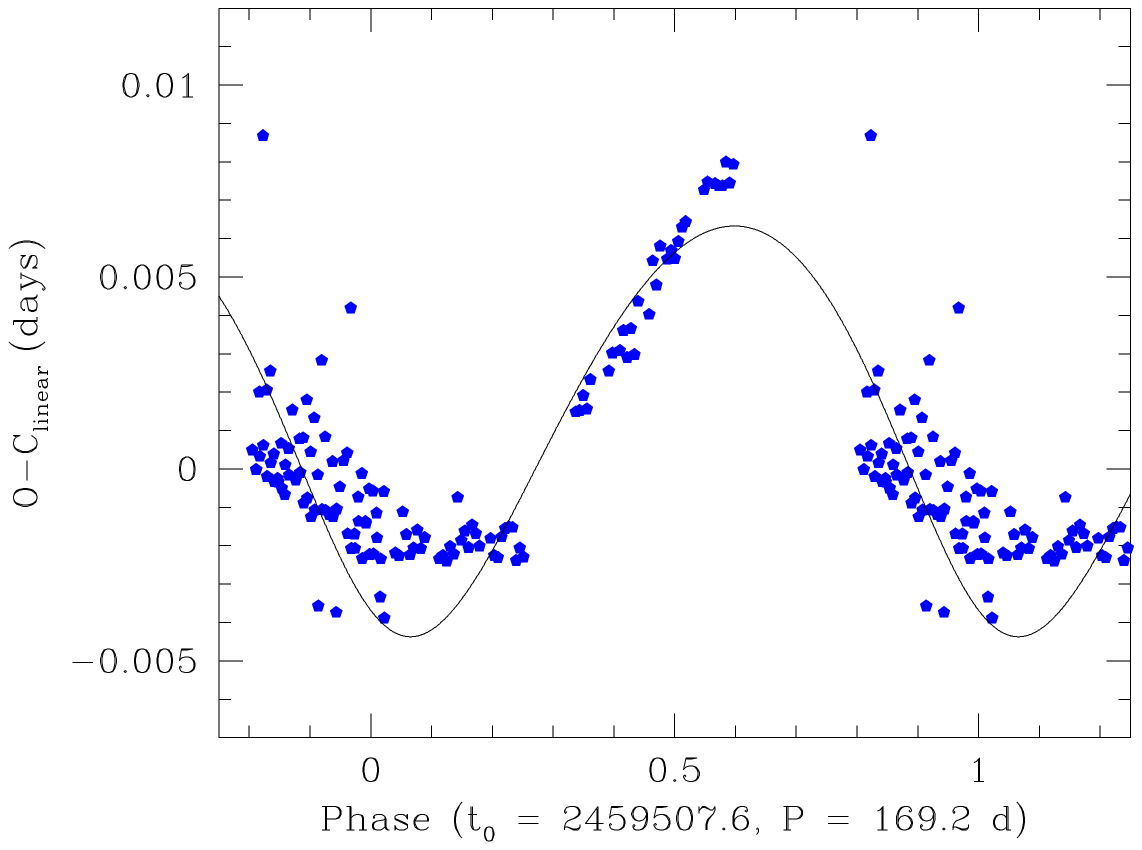}
  \end{center}  
\caption{Light-travel time effect adjustment of the $O-C$ values of the times of primary minimum of the inner eclipsing binary of TYC\,1881-933-1. Phase 0.0 corresponds to the periastron passage of the outer orbit.\label{OCltte}}
\end{figure*}

Finally, TYC\,1881-933-1 displays clear $O-C$ trends during each epoch, not only between observing campaigns (see the insets in the top left panel of Fig.\,\ref{SpOC}). More precise constraints can then be derived in this case because the $O-C$ needs to fold in a coherent way if it is due to LTTE. First, we determined the period of the LTTE effect in this target. The Fourier periodogram of the $O-C$ values of the times of primary minimum of TYC\,1881-933-1 displays a large number of peaks at frequencies below 0.01\,d$^{-1}$ (see the right panels of Fig.\,\ref{SpOC}). The frequencies of most of these peaks are integer multiples of the lowest frequency ($3.7\,10^{-4}$\,d$^{-1}$, corresponding to 2703\,d). Others are likely aliases. The complex spectral window has strong peaks (amplitude higher than 0.8) at $1.13\,10^{-3}$\,d$^{-1}$, $1.45\,10^{-3}$\,d$^{-1}$, $2.58\,10^{-3}$\,d$^{-1}$ and $4.03\,10^{-3}$\,d$^{-1}$, followed by somewhat lower peaks (amplitude above 0.65) at $2.90\,10^{-3}$\,d$^{-1}$ and $3.2\,10^{-4}$\,d$^{-1}$. 

We then adjusted the light-travel time relation of Eq.\,(\ref{ltte}) with the periods corresponding to each of these peaks and retained the periods that yielded the best adjustments to the data. They are 169.2, 225.2, 337.8 and 675.2\,d. The formally best-quality fit was obtained for the longest of these periods, although the rms of the fit is nearly independent of the period. The three shorter periods are the $n^{\rm th}$ harmonics of the 675\,d period, with $n = 2$, 3 and 4. Given the nonsinusoidal shape of the light-travel time relation, the 675\,d period thus appears to be the most promising candidate. The parameters for the four possible periods are given in Table\,\ref{solltte} and are displayed in Fig.\,\ref{OCltte}. Based on the periods and delays of that table and a binary mass of 7.3\,M$_{\odot}$ (Table \ref{binsol}), Kepler's law indicates a tertiary mass of $M_3=11-44$\,M$_{\odot}$ for $i=30-90^{\circ}$. The RV changes due to the gravity of the tertiary component would then amount to $K=a_{bin}\sin(i)\frac{2\pi}{P\sqrt{1-e^2}}\sim60$\kms. There are two problems with these evaluations. First, lines from a star that is this massive would be detectable in the spectra, which is not the case (only lines from the B2 primary are seen). Second, these strong RV changes would be seen while the RV curve remains similar, regardless of the epoch under consideration (LAMOST RVs were taken over 800\,d with eight campaigns with durations of 1--8\,d; see Fig. \ref{rvphot} and Table \ref{rv}; and they sample phases near apastron and near periastron). Again, more data are needed to clarify this result.

\begin{figure*}
  \begin{center}
    \includegraphics[width=6cm,bb=50 175 575 585, clip]{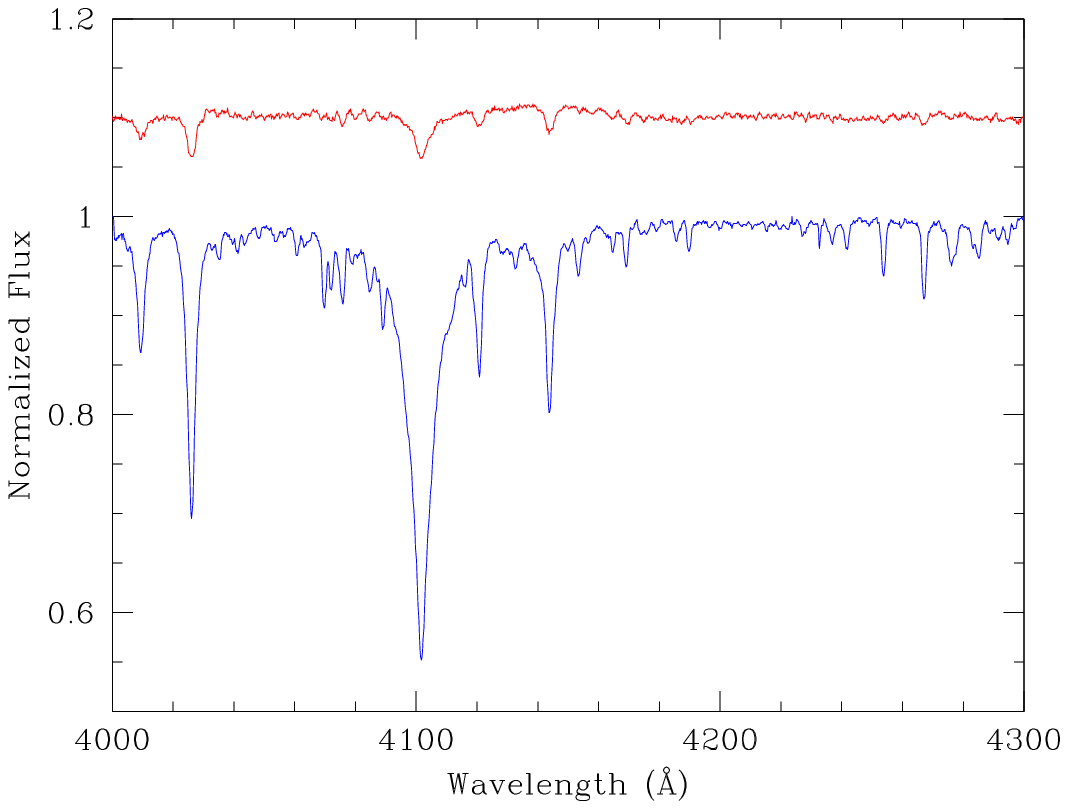}
    \includegraphics[width=6cm,bb=50 175 575 585, clip]{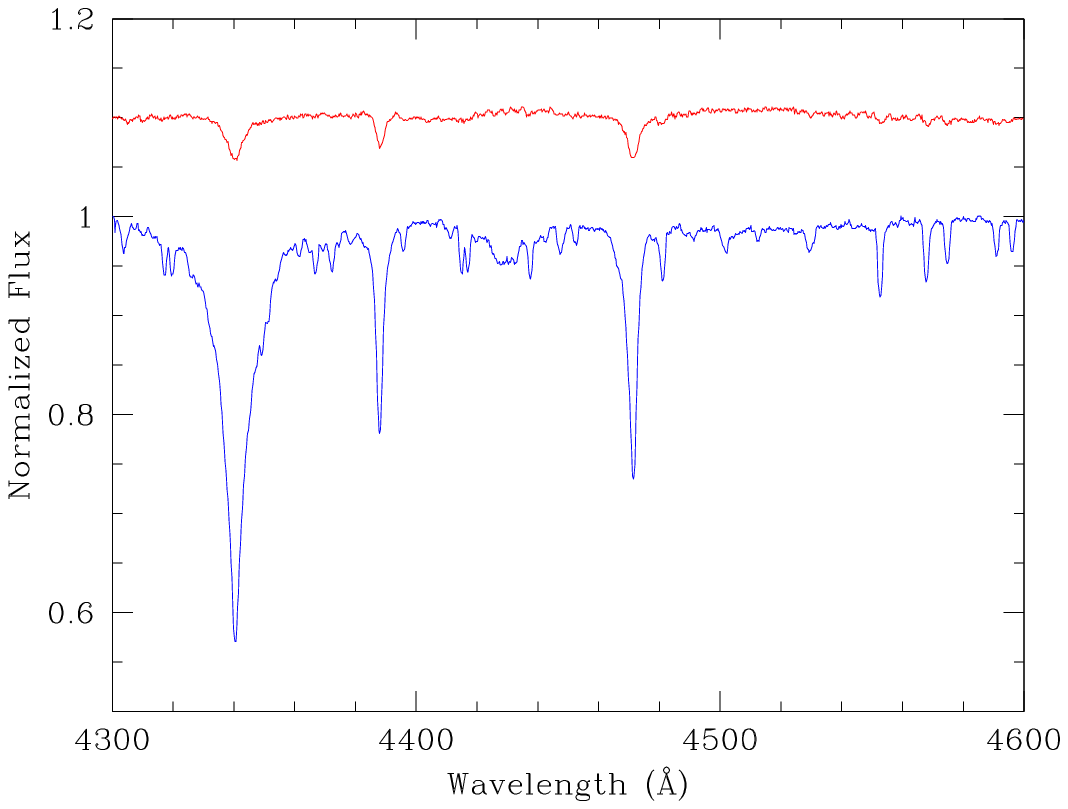}
    \includegraphics[width=6cm,bb=50 175 575 585, clip]{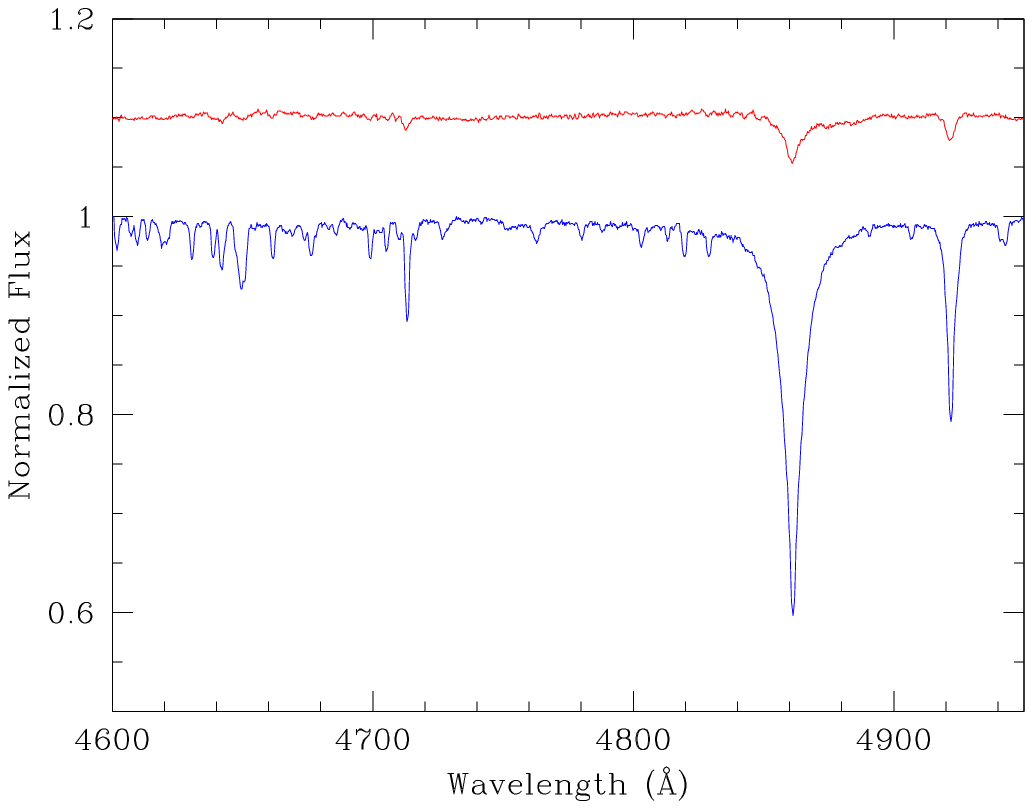}
  \end{center}  
  \caption{Result of the disentangling for V1061\,Cen. The main component is shown in blue, and the fainter component is shown in red (continuum shifted by +0.2 for clarity). No adjustment for the light ratio was made. The line depths are those observed in the input spectra, i.e., with respect to the combined light of all stars in the system. \label{v1061disent}}
\end{figure*}

\subsection{Spectral disentangling of V1061\,Cen}
For V1061\,Cen, the main spectral component appears to be stationary, while the fainter secondary component seems to shift with orbital phase. Fitting two Gaussians to the line profiles (one for the main component, the other for the secondary component) did not work well. Since lines appear in absorption, a first estimate of the spectra of the main component was built by taking the flux maximum of the five spectra at each wavelength step. This average spectrum was subtracted from all original spectra to isolate the secondary component, and a single Gaussian was then fit to the lines that appeared in this difference. Using these first estimates of the RVs, we performed a spectral disentangling on the whole spectral set using the shift-and-add procedure of \citet{gon06}, as coded by \citet{rau16}. This was done for three domains, 4000--4300\AA, 4300--4600\AA, and 4600--4950\AA. After fewer than 200 iterations, the results appeared to be stable. The final RVs, averaging the results from each of the three domains, are provided in Table \ref{rv} and shown in Fig. \ref{rvphot}. The spectral appearance (Fig. \ref{v1061disent}) suggests spectral types of B0V and B2.5V for the main and secondary components, respectively. As the main component is stationary, V1061\,Cen clearly is a triple system, but only one component of the eclipsing binary is detected. We attempted a disentangling with three components, but found no significant trace of the missing binary component. With an amplitude of $209.6\pm5.2$\,\kms\ and a period of 2.21\,d (Table \ref{phot1}), the mass function reaches $f(m)=2.11\pm0.16$\,M$_{\odot}$. Assuming the detected star has 6\,M$_{\odot}$ and considering $i\sim90^{\circ}$ because of the presence of eclipses, the mass of the other component should be around 7\,M$_{\odot}$, implying a star more massive than the component detected in the spectra. Since the RV of the detected component decreases around the main eclipse (Fig. \ref{rvphot}), the other component should also be cooler. Without a trace of this component in the spectra, however, we were prevented from deriving a full solution for this system. A more thorough investigation is now needed that probes several epochs with a better phase coverage.  

\section{Discussion and conclusions}

Our combined analysis of photometric and spectroscopic data adds a dozen systems in which a massive star is paired with a luminous companion of much lower mass. The system characteristics can be found in Tables \ref{targetlist} and \ref{binsol}. In all cases, this companion is cooler than the massive star, and it is therefore not expected to be a stripped star resulting from a binary interaction.

\begin{figure}
  \begin{center}
    \includegraphics[width=8cm]{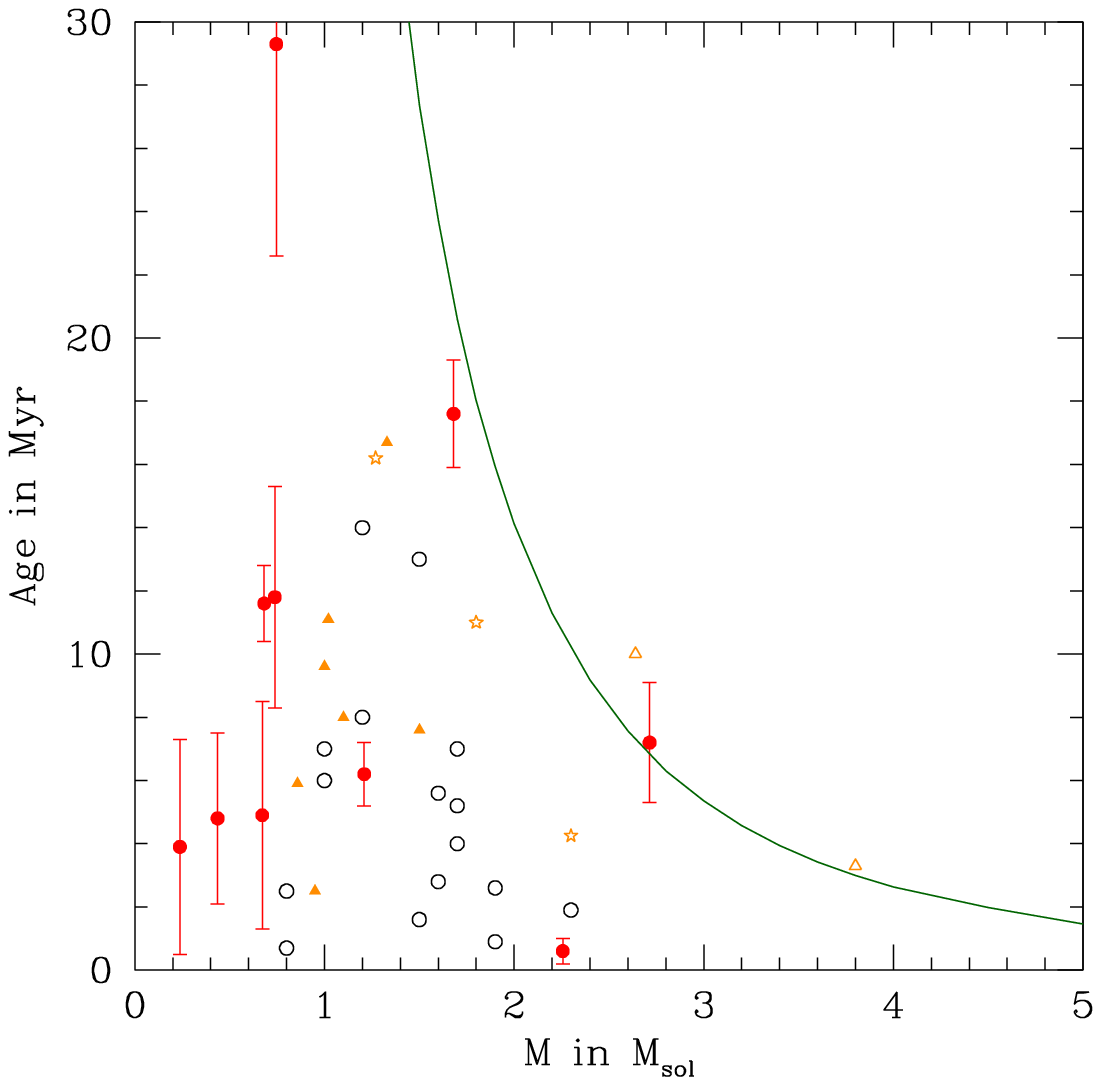}
  \end{center}  
  \caption{Comparison of mass and age for secondary stars of extreme mass ratio systems. Our targets are shown by red symbols, the MC systems of \citet{moe15} are shown with black circles, and the other Galactic cases are shown with orange triangles or stars (see text for details). The ZAMS track was defined from models by \citet{tog11}, considering a central hydrogen abundance $X_c$ equal to 99.5\% of the initial value (which corresponds to the stabilization of the total nuclear-based luminosity in these models). \label{agem}}
\end{figure}

The location of our targets in Hertzsprung-Russell diagrams and mass-radii diagrams are provided in Fig. \ref{hrd}. An additional plot provides a comparison between masses and ages for the secondaries alone because all primaries appear to be beyond the zero-age main sequence (ZAMS; Fig. \ref{agem}). The mass ratios of AN\,Dor and HD\,112485 are slightly higher than $q=0.2$, which means that they do not fulfill the criterion of an extreme mass ratio. In addition, considering their ages, their secondaries might be young MS stars with early-F and late-B spectral types, respectively, rather than PMS stars. The other secondaries, however, are clearly less massive and still in a PMS stage.

In Figures \ref{hrd} and \ref{agem}, our targets (shown by red dots) are compared to other young systems with an early-type primary (earlier than B3), a short period ($P<10$\,d), and an extreme mass ratio ($q<0.2$). This set comprises 16 of the nascent systems from \citet{moe15} in the Magellanic Clouds, shown by black circles, which were analyzed on the basis of photometry alone (spectroscopy is still required to confirm their status). In addition, this set also contains several Galactic cases whose analysis usually combined spectroscopy and photometry (Table \ref{whole}). Some of them (shown by open orange triangles in the figure) may not be true nascent binaries (i.e., systems with PMS secondaries), but rather young binaries with stars very close to the ZAMS: HD\,163892 \citep{mah22}, HD\,165246 \citep{joh21}, and HD\,210478 (V446\,Cep, \citealt{car14}). Other cases, shown by filled orange triangles, clearly harbor PMS secondaries, as demonstrated by their position in the age-mass diagram: HD\,145482 (c02\,Sco, \citealt{gul16,pig24}), HD\,149834 \citep{sta21}, HD\,120307, HD\,138690A  ($\nu$\,Cen and $\gamma$\,Lup\,A, \citealt{jer21}), HD\,25631, HD\,191495, and HD\,46485 \citep{naz23}. Three additional systems (shown by orange stars in the age-mass diagram) most probably also belong to this group, although only part of their physical parameters have been derived, and more study is required to fully constrain the full set of their properties: HD158926A ($\lambda$\,Sco\,A, \citealt{tan06}), HD\,152200 \citep{ban23,bri23}, and HD\,62747 (V390\,Pup, \citealt{pig24}). 

Binary systems, especially nascent ones, allow us to peer directly into the results of the stellar formation processes. As an example, \citet{smi24} used the {\it Gaia} DR3 catalog to study close ($d<500$\,pc) astrometric binaries with periods of 100--3000\,d and intermediate-mass primaries (spectral types between B8 and F1). They found that systems with the lower mass ratios ($q<0.32$) and/or lower eccentricities ($e<0.2$) have primaries that rotate more slowly and display a higher spin-orbit alignment than primaries in other systems. This excess of slow rotators supports the disk-fragmentation scenario, in which a very low-mass fragment is formed from a disk instability, then migrates inward, and finally accretes from the (now circumbinary) disk. In this case, higher accretion rates lead to more similar stars in more circular orbits, so that modest accretion is required to keep the mass ratio at low values. The lower eccentricity would then need to naturally arise from the fragmentation+migration process \citep{smi24}. The slow rotation of the primaries is then explained by two effects: (1) the circumbinary accretion mostly occurs on the secondary, so that after fragmentation, the primary star gains less in mass and angular momentum; (2) magnetic braking may transfer angular momentum from the primary through the (circumprimary) disk into the orbit. 

For systems with shorter periods, as studied here, the situation is expected to be quite different. The proximity of the two stars leads to a truncation of the circumprimary disk, prohibiting magnetic braking (\citealt{smi24} and references therein). Furthermore, because of their very low masses, the secondaries exert little tidal effects on the massive primaries \citep{bri24}, which limits braking of the rotation and therefore synchronization onto the orbital period. All this implies that stars born rotating fast will remain so. In this context, we compared the orbital period $P$ to the rotation period $P_{rot}$ of the primary derived from the projected rotational velocity, the radius, and the binary inclination. This revealed that only TYC\,1881-933-1 and LS\,I\,+61\,145 are synchronized. The rotation of other primary stars is clearly supersynchronous (Table \ref{binsol}): $P_{rot}/P\sim0.3$ for HD\,254346, HD\,112485, and HD\,350685 or 0.5--0.8 for AN\,Dor, TYC\,741-1565-1, HD\,309317, HD\,154407C, and V1208\,Sco. This fast rotation\footnote{It is however important to note that the rotation rates of our targets remain far from critical, $v_{rot}/v_{crit}\sim0.1-0.4$.} was previously found for the three systems that we analyzed in \citet{naz23}, and this also applies to the other nascent systems reported in the literature, except for HD\,120307 and HD\,62747 (Table \ref{whole} and Fig. \ref{histo}).

Supersynchronous rotation in nascent systems may appear to be unsurprising, but the high incidence of these cases in this group is clearly not expected. For example, \citet{hua10} studied rotation in B stars and demonstrated that the higher-mass ($M>8$\,M$_{\odot}$) cases rotate more slowly, and very fast rotation ($v/v_{crit}>0.7$, see their Fig. 6) is even totally absent for them. One may wonder whether the observed large fraction of supersynchronous rotators among the nascent short-period binaries with extreme mass ratios is an observational bias. A priori, stellar rotation plays no role in the appearance of eclipses and reflection effects that were used to detect them. Observations may thus reflect the actual situation. In support of this idea, we may note that the vast majority of young HAeBe stars analyzed by \citet{pin21} are not slow rotators, but display $v\sin(i)>100$\,\kms. It is true that the sample of nascent systems currently remains limited, however, with only 18 cases. Increasing the sample would certainly improve the reliability of these conclusions and clarify the rotation distribution. In this context, we may note that disk fragmentation modeling and observations revealed the largest fraction of systems with very low mass ratios to lie in long-period ($P>50$\,d) binaries \citep{tok20}. While long periods make the spectroscopic detection of the orbital motion of the primaries even more difficult than for the cases studied here, the study of these systems would also help us to reach a broader view of this population, and it would thereby further constrain the star formation scenarios.

\section{Data availability}
Figures showing all light curves are available on Zenodo (https://doi.org/10.5281/zenodo.17257979).

\begin{acknowledgements}
The Li\`ege authors acknowledge support from the Fonds National de la Recherche Scientifique (Belgium) and the University of Li\`ege (through its Special Funds for Research and the IPD-STEMA program). This paper includes data collected by the \te\ mission, which are publicly available from the Mikulski Archive for Space Telescopes (MAST). Funding for the \te\ mission is provided by NASA's Science Mission directorate. ADS and CDS were used for preparing this document. This project has received funding from the European Union's Horizon 2020 research and innovation programme under grant agreement No 101004719 (OPTICON/ORP) and J.L.-B. was co-funded by the European Union (ERC, MAGNIFY, Project 101126182) - this material reflects only the authors views and the Commission is not liable for any use that may be made of the information contained therein. 
\end{acknowledgements}

\bibliographystyle{aa}
\bibliography{binaries}

\appendix
\section{Tables presenting the target and their properties}

\begin{sidewaystable}
  \begin{center}
  \scriptsize
  \caption{Targets and their basic properties (see text for details), ordered by increasing RA.
 \label{targetlist}}
  \begin{tabular}{lcccccccccccc}
    \hline
Name & Sp. Type & $d$(pc) & $E(B-V)$ & $\log(L_{\rm BOL}/L_{\odot})$ & $T_{\rm eff}^1$(K) & $\log(g^1)$ & $v^1\,\sin(i)$(\kms) & $M_{evol}^1$(M$_{\odot}$) & $R_{evol}^1$(R$_{\odot}$) & Age (Myr) & Known EB? & Sectors \\
\hline
LS\,I\,+61\,145	&B1V	& 2625$\pm$79 & 0.46 & 3.97$\pm$0.06 & 26000 (fixed) &               & 109.5$\pm$1.3 & 11.2$\pm$0.7 & 5.0$\pm$0.5 &  4.9$\pm$3.6 & g,i,j,p                 & 17,18,{\bf 58*,78*,85*} \\
LS\,I\,+61\,275	&B1.5V	& 2142$\pm$85 & 0.62 & 3.74$\pm$0.08 & 24630$\pm$388 &               & 160$\pm$12    &  9.6$\pm$0.6 & 4.1$\pm$0.4 &  3.3$\pm$3.0 & a,h,j,p                 & {\bf 18,58,85,86} \\
AN\,Dor	        &B2V	&  950$\pm$38 & 0.02 & 3.65$\pm$0.08 & 21529$\pm$65  & 4.02$\pm$0.01 & 162.1$\pm$1.1 &  8.2$\pm$0.3 & 4.7$\pm$0.1 & 17.6$\pm$1.7 & c,f,g,i,j,k,n,p,r,s,t,u & $\dagger$ \\
TYC\,1881-933-1	&B2V	& 1928$\pm$96 & 0.33 & 3.40$\pm$0.10 & 18700$\pm$428 & 4.11$\pm$0.12 & 143.3$\pm$20.7&  6.6$\pm$0.4 & 4.1$\pm$0.5 & 29.3$\pm$6.7 & b,m,p                   & {\bf K} + 43,44,45,71,72 \\
HD\,254346	&B1.5V	& 1730$\pm$82 & 0.45 & 4.02$\pm$0.09 & 24757$\pm$145 & 3.80$\pm$0.02 & 201.2$\pm$1.9 & 11.0$\pm$0.5 & 5.9$\pm$0.5 & 11.6$\pm$1.2 & g,l                     & 43*,{\bf 44*,45*,71*,72*} \\
TYC\,741-1565-1	&B2V	& 3030$\pm$154& 0.50 & 3.49$\pm$0.10 & 23634$\pm$391 & 3.95$\pm$0.05 & 148.9$\pm$5.8 &  8.8$\pm$0.3 & 3.8$\pm$0.3 &  3.9$\pm$3.4 & h,s,t                   & {\bf 6*,} 33, {\bf 71*,72*} \\
HD\,309317	&B2V	& 2197$\pm$95 & 0.21 & 3.80$\pm$0.09 & 22978$\pm$167 & 4.14$\pm$0.02 & 203.1$\pm$2.6 &  9.0$\pm$0.5 & 4.5$\pm$0.5 & 11.8$\pm$3.5 & i,j,v                   & 10,11,37,38,{\bf 64*}\\
HD\,112485	&B1.5V	& 1749$\pm$67 & 0.37 & 4.07$\pm$0.08 & 26096$\pm$659 & 4.06$\pm$0.07 & 184.0$\pm$8.4 & 11.6$\pm$0.6 & 5.2$\pm$0.4 &  7.2$\pm$1.9 & g,i,j                   & 11,38,{\bf 64*,65*}\\
V1061\,Cen	&B2V	& 1374$\pm$39 & 0.33 & 3.57$\pm$0.06 &               &               &               &              &             &              & c,g,p,q                 & 11,38,{\bf 65*} \\
V1208\,Sco      &B0.5V  & 1504$\pm$32 & 0.49 & 4.14$\pm$0.04 & 29600$\pm$100 & 4.25$\pm$0.02 &    87$\pm$8   & 13.6$\pm$0.2 & 4.6$\pm$0.1 &  0.6$\pm$0.4 & d,e,o                   & {\bf 12,39} \\
HD\,154407C	&B1V	& 1298$\pm$70 & 0.45 & 4.13$\pm$0.11 & 25696$\pm$467 & 4.32$\pm$0.06 & 144.8$\pm$7.3 & 11.0$\pm$0.6 & 4.6$\pm$0.4 &  4.8$\pm$2.7 & g,i,j                     & 12,{\bf 66*}\\
HD\,350685	&B1.5V	& 1974$\pm$82 & 0.37 & 3.98$\pm$0.08 & 25832$\pm$142 & 4.09$\pm$0.02 & 247.7$\pm$2.0 & 11.2$\pm$0.6 & 5.0$\pm$0.2 &  6.2$\pm$1.0 & h,i,j,v                 & 14,{\bf 41,54} \\
\hline
  \end{tabular}
  \end{center}
\tablefoot{Error on the bolometric luminosity is derived from the error on distance. The effective temperature of LS\,I\,+61\,145 could not be evaluated from spectra hence is assumed to be that corresponding to its spectral type. For LS\,I\,+61\,275, the effective temperature comes from \citet{hua06} and the projected rotational velocity from \citet{hua06b}. For TYC\,1881-933-1, the effective temperature, gravity, and projected rotation rates are averages from values reported for the star in \citet{xia22} and errors correspond to dispersions. For V1208\,Sco, the effective temperature, gravity, and projected rotation rates come from \citet{mcs09}. For the best-fit evolutionary parameters, the largest error is used if asymmetric. References for the eclipsing binary (EB) status are: (a) \citet{ant08}, (b) \citet{arm15}, (c) \citet{avv13}, (d) \citet{bal85}, (e) \citet{ban23}, (f) \citet{dub11}, (g) \citet{eze24}, (h) \citet{gre23}, (i) \citet{ijs21}, (j) \citet{ijs24}, (k) \citet{kaz99}, (l) \citet{lab20}, (m) \citet{lac15}, (n) \citet{mal06}, (o) \citet{mar24}, (p) \citet{mow23}, (q) \citet{ote03}, (r) \citet{per00}, (s) \citet{prs22}, (t) \citet{shi22}, (u) \citet{sou22}, and (v) \citet{tka24}. For the last column, $K$ indicates {\it Kepler} data, the asterisk $*$ identifies 2-min light curves, bold shows which Sectors were averaged and combined to get the binned light curve which was analyzed, and the $\dagger$ symbol corresponds to Sectors 2*,3*,4*,5*,6*,9,12,13,29*,30*,31*,32*,33*,36*,39*,{\bf 63*,64*,65*,} 66*,67*,69*. }
\end{sidewaystable}
 
\begin{sidewaystable}
  \begin{center}
  \scriptsize
  \caption{Basic properties derived from light curves.
 \label{phot1}}
  \begin{tabular}{lccccccccc}
    \hline
Name & $T_0$ (HJD$-2\,450\,000)$ & $f_{orb}$ (d$^{-1}$) &  $\langle I \rangle$ (mmag) & $\Delta I_1$ (mmag) & $\theta_1$ & $\Delta I_2$ (mmag) & $\phi_2$ & $\theta_2$ & $\Delta I_{refl}$  \\
\hline
LS\,I\,+61\,145  &10636.210$\pm$0.001 &0.541614$\pm$0.000001  &    1.4$\pm$0.8 & 86.0$\pm$1.7 & 0.0334$\pm$0.0008 &  30.9$\pm$1.6 &0.5012$\pm$0.0019 &0.0338$\pm$0.0023 & 22.7$\pm$1.5 \\
LS\,I\,+61\,275  &10635.563$\pm$0.001 &0.7569565$\pm$0.0000005& --14.9$\pm$2.3 &659.2$\pm$3.8 & 0.0397$\pm$0.0003 & 203.5$\pm$3.7 &0.4999$\pm$0.0008 &0.0460$\pm$0.0011 &150.6$\pm$4.3 \\
AN\,Dor          &10204.570$\pm$0.001 &0.4919645$\pm$0.0000005&  --4.2$\pm$0.8 &169.2$\pm$1.8 & 0.0287$\pm$0.0004 &  65.1$\pm$1.8 &0.5310$\pm$0.0009 &0.0298$\pm$0.0010 & 24.6$\pm$1.4 \\  
TYC\,1881-933-1  &10285.346$\pm$0.001 &0.980670$\pm$0.000001  &    1.3$\pm$0.9 &108.2$\pm$1.4 & 0.0426$\pm$0.0007 &  56.1$\pm$1.3 &0.5008$\pm$0.0011 &0.0498$\pm$0.0015 & 26.7$\pm$1.7 \\
HD\,254346 	 &10282.380$\pm$0.001 &0.184102$\pm$0.000001  &    0.5$\pm$0.1 & 17.1$\pm$0.2 & 0.0106$\pm$0.0001 &   6.7$\pm$0.2 &0.4939$\pm$0.0003 &0.0107$\pm$0.0003 &  2.4$\pm$0.1 \\
TYC\,741-1565-1  &10284.733$\pm$0.001 &0.443032$\pm$0.000001  &    0.6$\pm$0.2 & 26.3$\pm$0.5 & 0.0267$\pm$0.0006 &   8.5$\pm$0.5 &0.4999$\pm$0.0018 &0.0269$\pm$0.0020 &  5.9$\pm$0.4 \\
HD\,309317 	 &10066.796$\pm$0.001 &0.442999$\pm$0.000001  &    2.8$\pm$0.4 & 73.6$\pm$0.9 & 0.0267$\pm$0.0004 &  24.8$\pm$0.9 &0.5007$\pm$0.0011 &0.0276$\pm$0.0012 & 18.4$\pm$0.6 \\
HD\,112485 	 &10093.109$\pm$0.001 &0.186135$\pm$0.000001  &    0.3$\pm$0.2 & 61.5$\pm$0.7 & 0.0129$\pm$0.0002 &  26.2$\pm$0.7 &0.5000$\pm$0.0004 &0.0137$\pm$0.0004 &  6.3$\pm$0.3 \\
V1061\,Cen 	 &10094.351$\pm$0.001 &0.452576$\pm$0.000001  &  --8.1$\pm$0.4 &141.6$\pm$1.1 & 0.0247$\pm$0.0002 &  57.8$\pm$1.1 &0.4948$\pm$0.0006 &0.0260$\pm$0.0006 &  9.5$\pm$0.7 \\
V1208\,Sco       & 9382.280$\pm$0.001 &0.191583$\pm$0.000001  &  --6.7$\pm$0.9 &186.3$\pm$3.0 & 0.0175$\pm$0.0003 &  52.2$\pm$2.9 &0.4999$\pm$0.0011 &0.0181$\pm$0.0012 &  8.6$\pm$1.5\\
HD\,154407C 	 &10118.219$\pm$0.001 &0.390805$\pm$0.000005  &  --4.2$\pm$0.2 & 76.9$\pm$0.7 & 0.0217$\pm$0.0002 &  34.2$\pm$0.7 &0.5003$\pm$0.0005 &0.0226$\pm$0.0006 &  4.3$\pm$0.4 \\
HD\,350685 	 & 9796.517$\pm$0.001 &0.347972$\pm$0.000001  &   24.1$\pm$0.4 & 94.5$\pm$1.5 & 0.0166$\pm$0.0003 &  24.7$\pm$1.6 &0.5005$\pm$0.0011 &0.0141$\pm$0.0011 & 58.2$\pm$0.7 \\
\hline
AN\,Dor (ASAS2)  &                    &                       & 7.682$\pm$0.002&169$\pm$4     & 0.0281$\pm$0.0009 &  55.6$\pm$4.6 &0.4800$\pm$0.0033 &0.0376$\pm$0.0043 & 27.6$\pm$3.8 \\
AN\,Dor (ASAS3)  &                    &                       & 7.676$\pm$0.003&185$\pm$10    & 0.0249$\pm$0.0016 &  58.9$\pm$6.1 &0.4687$\pm$0.0036 &0.0331$\pm$0.0044 & 24.9$\pm$5.0 \\
AN\,Dor (ASAS4)  &                    &                       & 7.671$\pm$0.003&158$\pm$9     & 0.0281$\pm$0.0016 &  55.0$\pm$6.8 &0.4831$\pm$0.0036 &0.0311$\pm$0.0042 & 14.2$\pm$5.3 \\
\hline
  \end{tabular}
  \end{center}
  \tablefoot{Reference times $T_0$ and phases $\phi=0$ correspond to the deeper, main eclipses, while $P=1/f_{orb}$ is the orbital period and the other columns yield the parameters of the Moe \& Di Stefano model (see text for details). For {\it ASAS} data, $\langle I \rangle$ is given in mag, not mmag, and intervals 2, 3, and 4 correspond to HJD$-2\,450\,000=$ 2350--3200, 3200--4000, and $>4000$, respectively.  }
\end{sidewaystable}

\begin{sidewaystable}
  \scriptsize
  \caption{Light curve and orbital solutions for our targets. \label{binsol}}
  \begin{center}
  \begin{tabular}{lcccccccccc}
    \hline
Parameter & LS\,I\,+61\,145 & AN\,Dor & TYC\,1881-933-1 & HD\,254346 & TYC\,741-1565-1 & HD\,309317 & HD\,112485 & HD\,350685 & HD\,154407C & V1208\,Sco\\
\hline
$M_1$(M$_{\odot}$)-fixed & 11.2 & 8.2 & 6.6 & 11 & 8.8 & 9 & 11.6 & 11.2 & 11 & 13.12 \\
$T_{\rm eff}^1$(K)-fixed & 26000 & 21529 & 18700 & 24757 & 23634 & 22978 & 26096 & 25832 & 25696 & 29600 \\
$P$(d)-fixed      & 1.846333 & 2.032883 & 1.019711 & 5.431772 & 2.257173 & 2.257341 & 5.372445 & 2.873794 & 2.558821 & 5.219670  \\
$\gamma$(\kms)         & --36.5$\pm$1.9 & 54.9$\pm$2.0 & 15.1$\pm$0.4 & 11.9$\pm$2.5 & 44.3$\pm$1.6 & --10.3$\pm$1.4 & --19.3$\pm$2.2 & 20.8$\pm$3.0 & 20.3$\pm$1.9 & --26.5$\pm$ 0.3 \\
$K_1$(\kms)            &   22.2$\pm$2.8 & 59.6$\pm$2.8 & 38.7$\pm$0.5 & 15.6$\pm$3.8 &  8.7$\pm$2.2 &   25.6$\pm$2.0 &   55.0$\pm$2.8 & 32.3$\pm$5.3 & 13.2$\pm$2.2 & 44.7$\pm$0.4 \\
$f(m)$ (M$_{\odot}$)    & 0.00209$\pm$0.00079 & 0.0444$\pm$0.0063 & 0.00612$\pm$0.00024 & 0.00213$\pm$0.00155 & 0.000153$\pm$0.000117 & 0.00393$\pm$0.00093 & 0.0923$\pm$0.0139 & 0.0100$\pm$0.0049 & 0.00060$\pm$0.00035 & 0.0483$\pm$0.0013\\
$e$-fixed             & 0 & 0.052 & 0 & 0.015 & 0 & 0 & 0 & 0 & 0 & 0\\
$\omega$($^{\circ}$)   &   & 340.76(fixed) &   & 130.8$\pm$1.0\\					
$i$($^{\circ}$)        & 81.5$\pm$0.05 & 75.90$\pm$0.06 & 68.07$\pm$0.20 & 76.74$\pm$0.10 & 73.17$\pm$0.10 & 77.60$\pm$0.10 & 78.53$\pm$0.15 & 72.70$\pm$0.25 & 79.14$\pm$0.20 & 85.37$\pm$0.20\\
$q$-fixed              & 0.060 & 0.205 & 0.113 & 0.062 & 0.027 & 0.082 & 0.234 & 0.108 & 0.040 & 0.172\\
$M_2$(M$_{\odot}$)-fixed & 0.67 & 1.68 & 0.75 & 0.68 & 0.24 & 0.74 & 2.71 & 1.21 & 0.44 & 2.26 \\
$f_1$ & 0.569$\pm$0.004 & 0.676$\pm$0.005 & 0.792$\pm$0.003 & 0.416$\pm$0.001 & 0.602$\pm$0.001 & 0.561$\pm$0.001 & 0.412$\pm$0.003 & 0.458$\pm$0.007 & 0.454$\pm$0.007 & 0.359$\pm$0.003 \\
$f_2$ & 0.517$\pm$0.001 & 0.601$\pm$0.005 & 0.780$\pm$0.002 & 0.239$\pm$0.001 & 0.422$\pm$0.001 & 0.431$\pm$0.003 & 0.289$\pm$0.011 & 0.751$\pm$0.006 & 0.463$\pm$0.010 & 0.300$\pm$0.005 \\
$R_1$(R$_{\odot}$) & 4.631$\pm$0.032 & 4.521$\pm$0.032 & 3.468$\pm$0.014 & 6.803$\pm$0.015 & 5.441$\pm$0.008 & 4.742$\pm$0.008 & 6.141$\pm$0.042 & 4.782$\pm$0.070 & 4.704$\pm$0.073 & 5.632$\pm$0.047 \\
$R_2$(R$_{\odot}$) & 1.221$\pm$0.002 & 1.948$\pm$0.016 & 1.293$\pm$0.003 & 1.140$\pm$0.005 & 0.808$\pm$0.002 & 1.195$\pm$0.008 & 2.206$\pm$0.080 & 2.912$\pm$0.023 & 1.178$\pm$0.025 & 2.108$\pm$0.035 \\
$T_{\rm eff}^2$(K) & 8082-10286$\pm$500 & 9071-9941$\pm$150 & 8671-9826$\pm$120 & 11680-12327$\pm$50 & 8295-10319$\pm$175 & 7777-9430$\pm$150 & 13625-14017$\pm$100 & 8074-9536$\pm$150 & 14214-14784$\pm$300 & 12252-12951$\pm$400 \\
$P_{rot}(d)$ & 2.12$\pm$0.03 &1.37$\pm$0.01 &1.14$\pm$0.17 &1.67$\pm$0.02 &1.77$\pm$0.07 &1.15$\pm$0.01 &1.66$\pm$0.08 &0.93$\pm$0.02 &1.61$\pm$0.09 &3.27$\pm$0.30 \\
$\log(g^1)$ & 4.16$\pm$0.01 & 4.04$\pm$0.01 & 4.18$\pm$0.01 & 3.81$\pm$0.01 & 3.91$\pm$0.01 & 4.04$\pm$0.01 & 3.93$\pm$0.01 & 4.13$\pm$0.01 & 4.43$\pm$ & 4.05$\pm$0.01 \\
$\log(g^2)$ & 4.09$\pm$0.01 & 4.08$\pm$0.01 & 4.09$\pm$0.01 & 4.16$\pm$0.01 & 4.00$\pm$0.01 & 4.15$\pm$0.01 & 4.18$\pm$0.03 & 3.59$\pm$0.01 & 3.94$\pm$0.02 & 4.14$\pm$0.01 \\
    \hline
  \end{tabular}
  \end{center}
  \tablefoot{The RV solution for AN\,Dor comes from Table \ref{omegadot_photo} and that of V1208\,Sco from \citet{ban22} - see text for details. For the other targets, the zero point $\gamma$ and amplitude $K_1$ were calculated assuming zero eccentricity with the period and $T_0$ from Table \ref{phot1}. To calculate the photometric solutions of all targets, we fixed the eccentricities and ephemerides (Table \ref{phot1}), primary masses and primary effective temperatures (from Table \ref{targetlist}), and mass ratios (derived from the mass functions $f(m)$ and a preliminary fitting to get inclinations $i$). For secondary temperatures, the first value provides the usual effective temperature (i.e., without any reflection effect) while the second value corresponds to a mean over the stellar surface, that is, it accounts for the heating of some parts of the star because of the reflection effect. The rotation period $P_{rot}$ is calculated using $2\pi R_1 \sin(i)/v \sin(i)$. }
\end{sidewaystable}

\section{Additional information}

\begin{figure}
  \begin{center}
    \includegraphics[width=8cm]{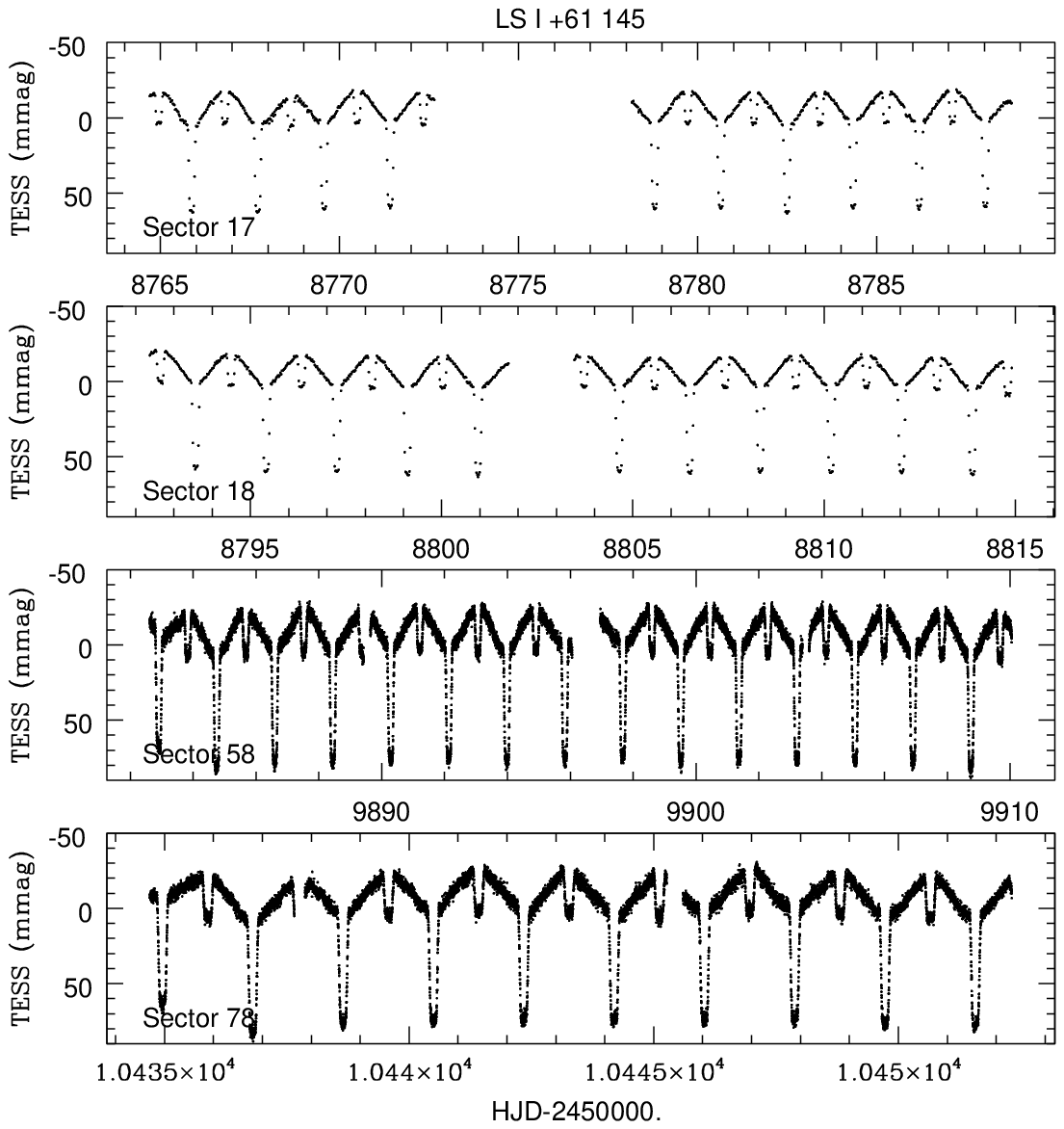}
  \end{center}  
  \caption{Individual light curves of the targets - sample. The full set is provided as an additional pdf on Zenodo. \label{LC}}
\end{figure}

\begin{figure*}
  \begin{center}
    \includegraphics[width=5.5cm]{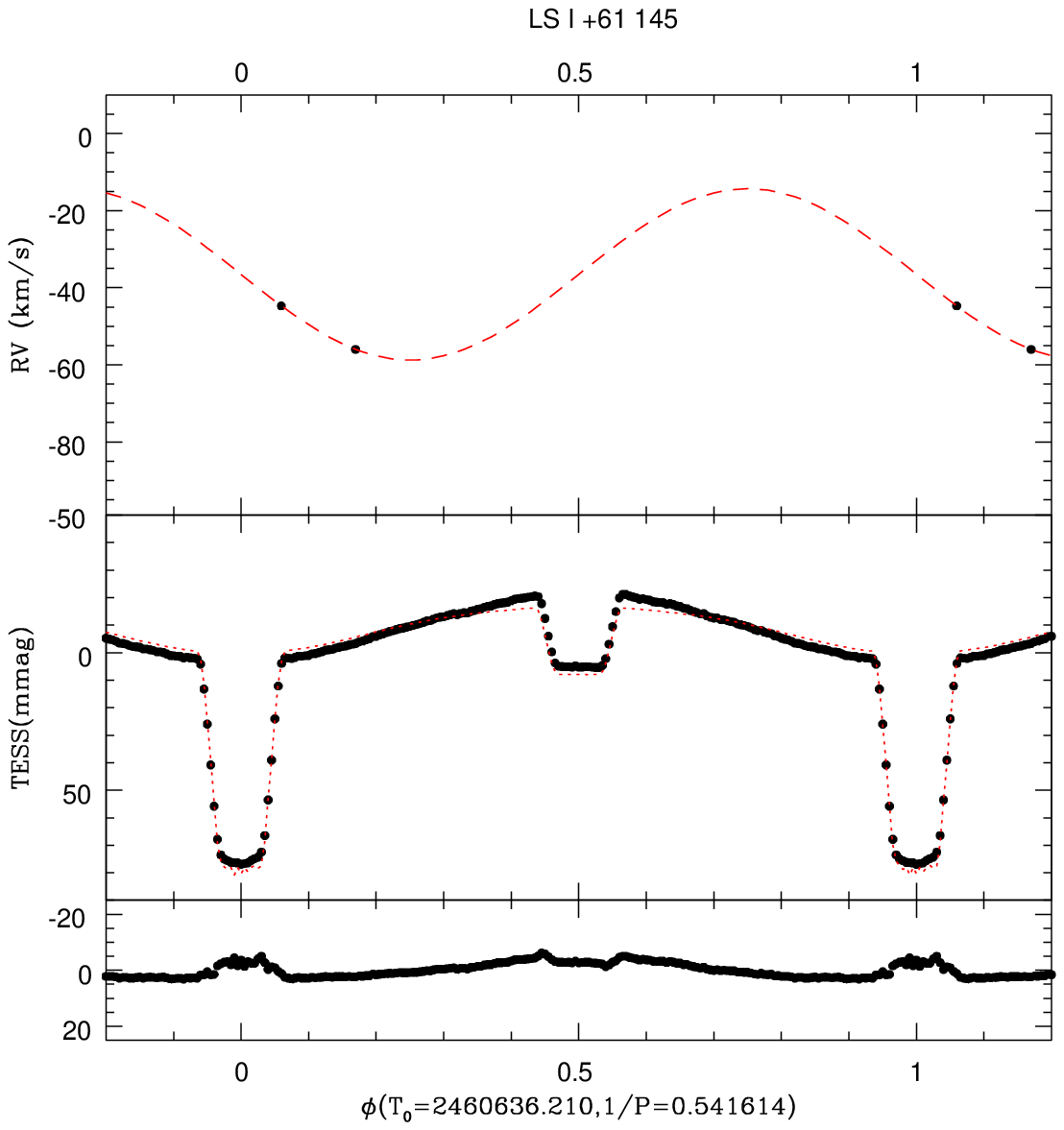}
    \includegraphics[width=5.5cm]{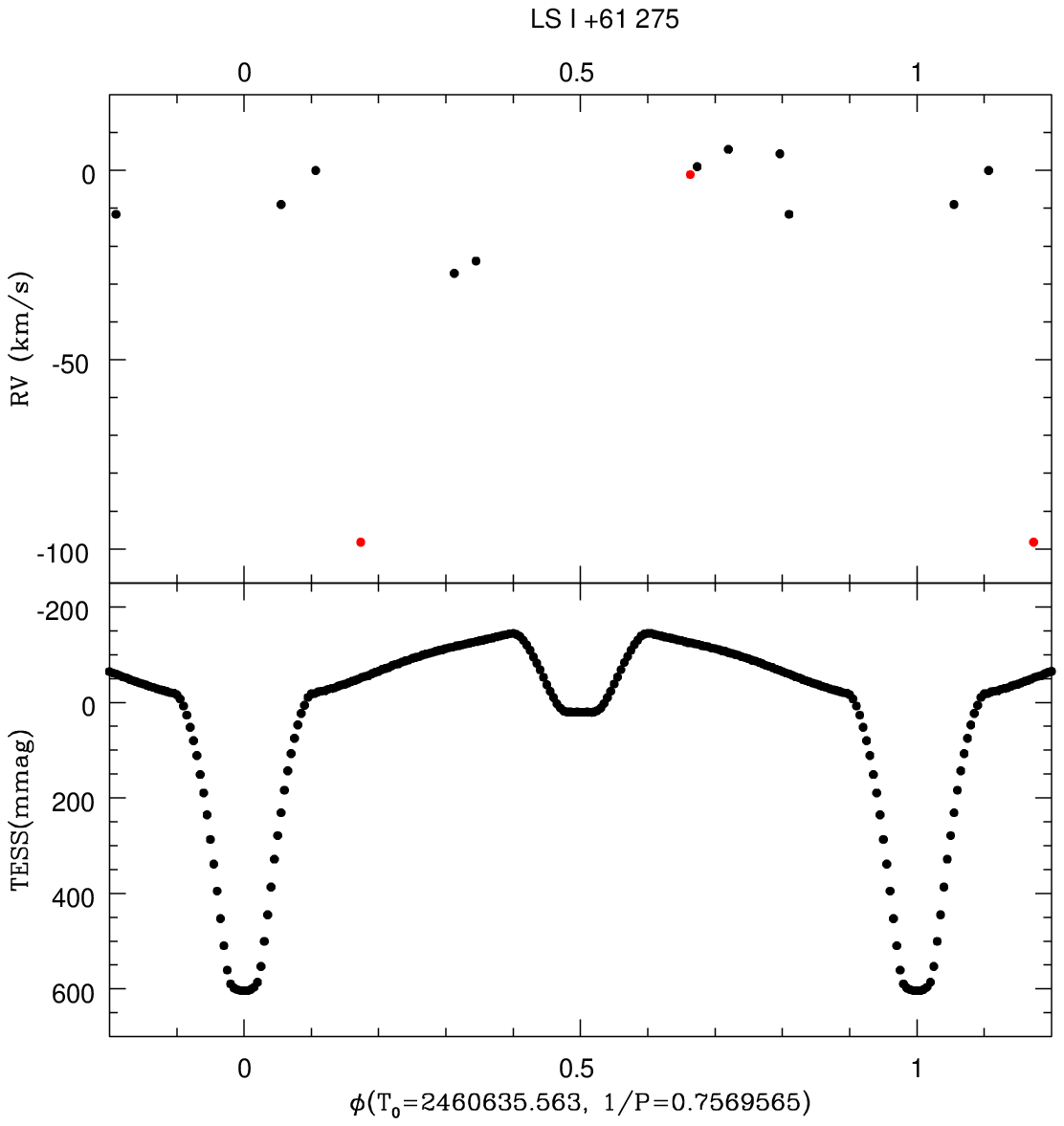}
    \includegraphics[width=5.5cm]{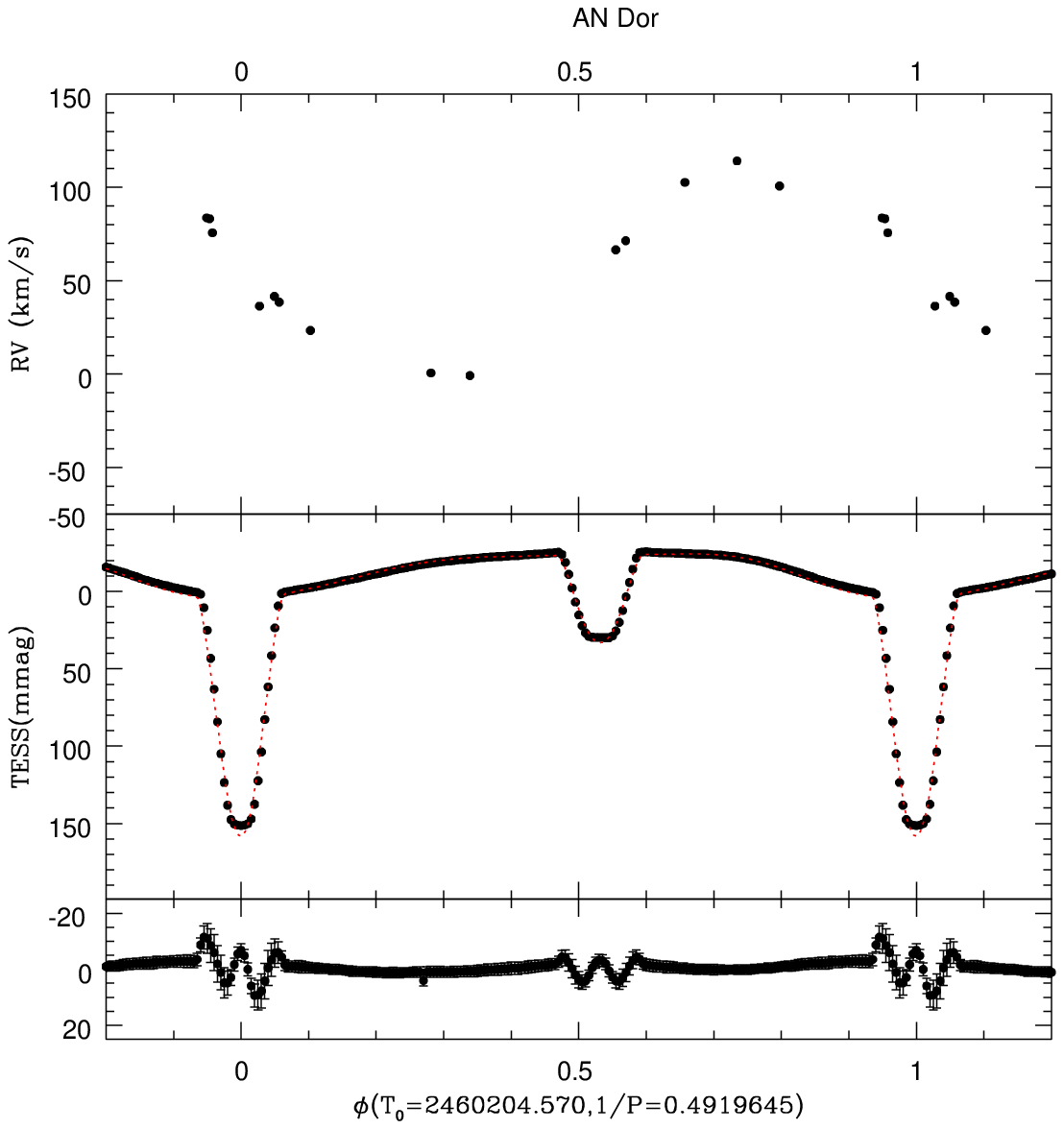}
    \includegraphics[width=5.5cm]{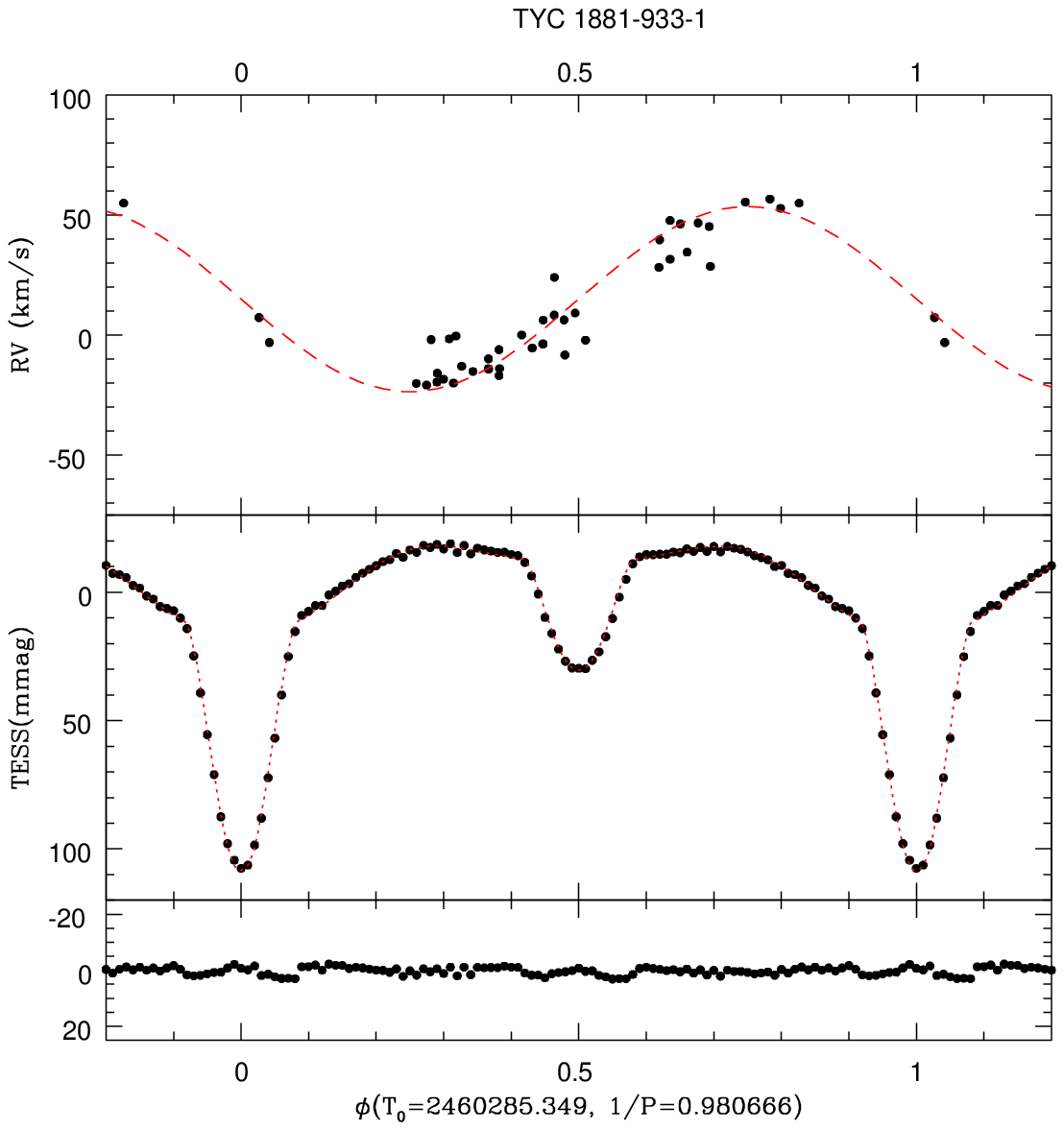}
    \includegraphics[width=5.5cm]{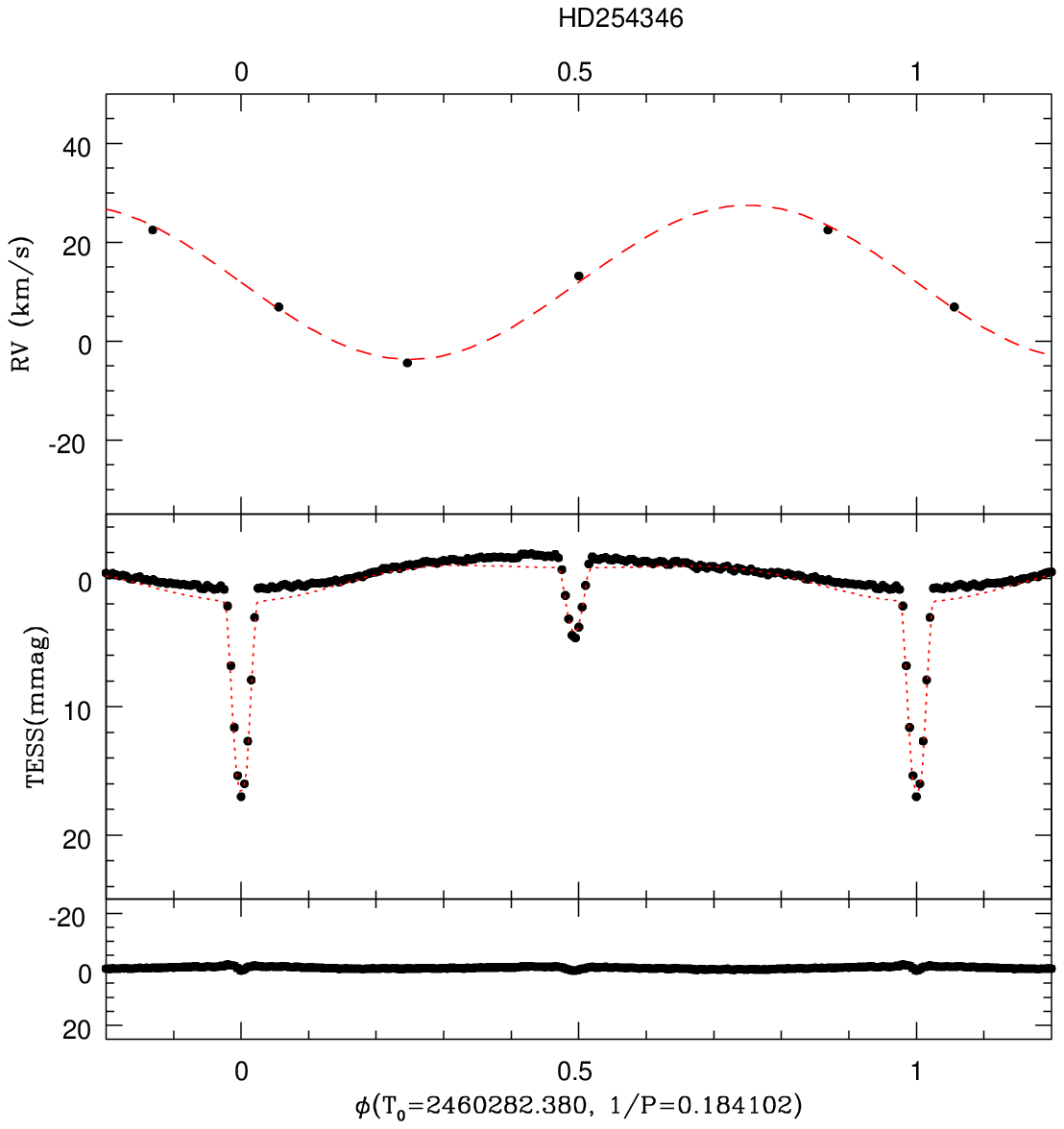}
    \includegraphics[width=5.5cm]{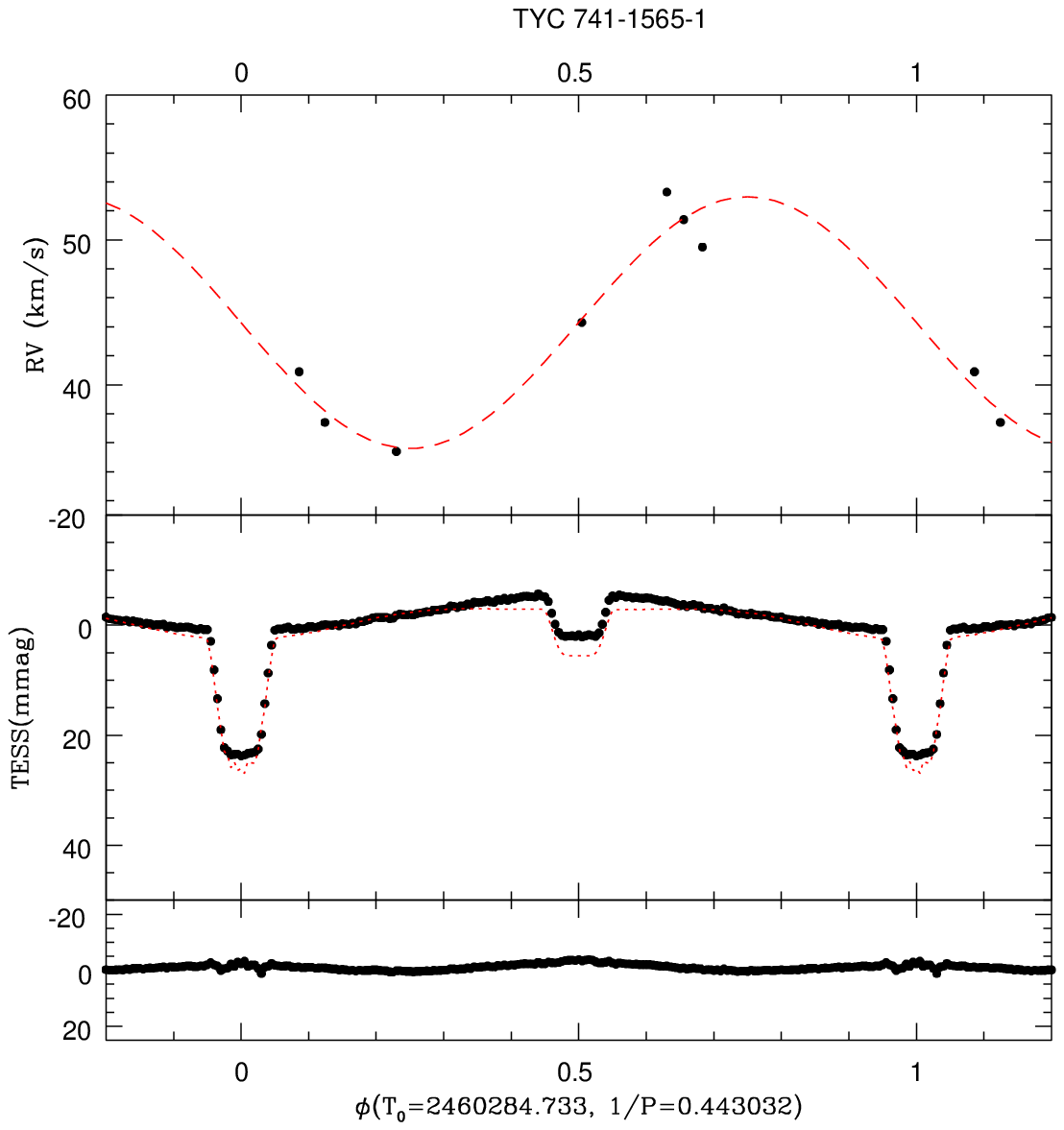}
    \includegraphics[width=5.5cm]{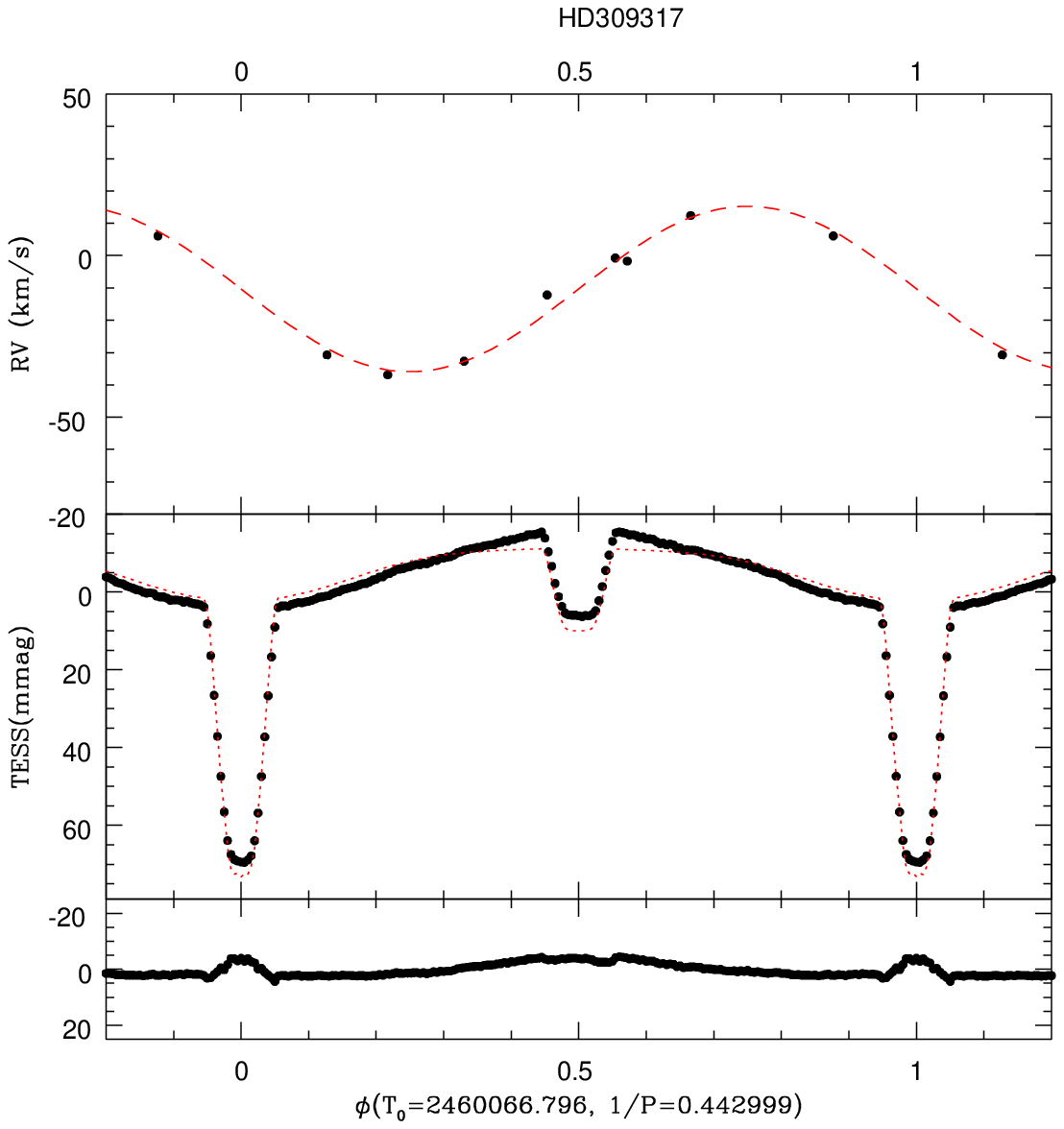}
    \includegraphics[width=5.5cm]{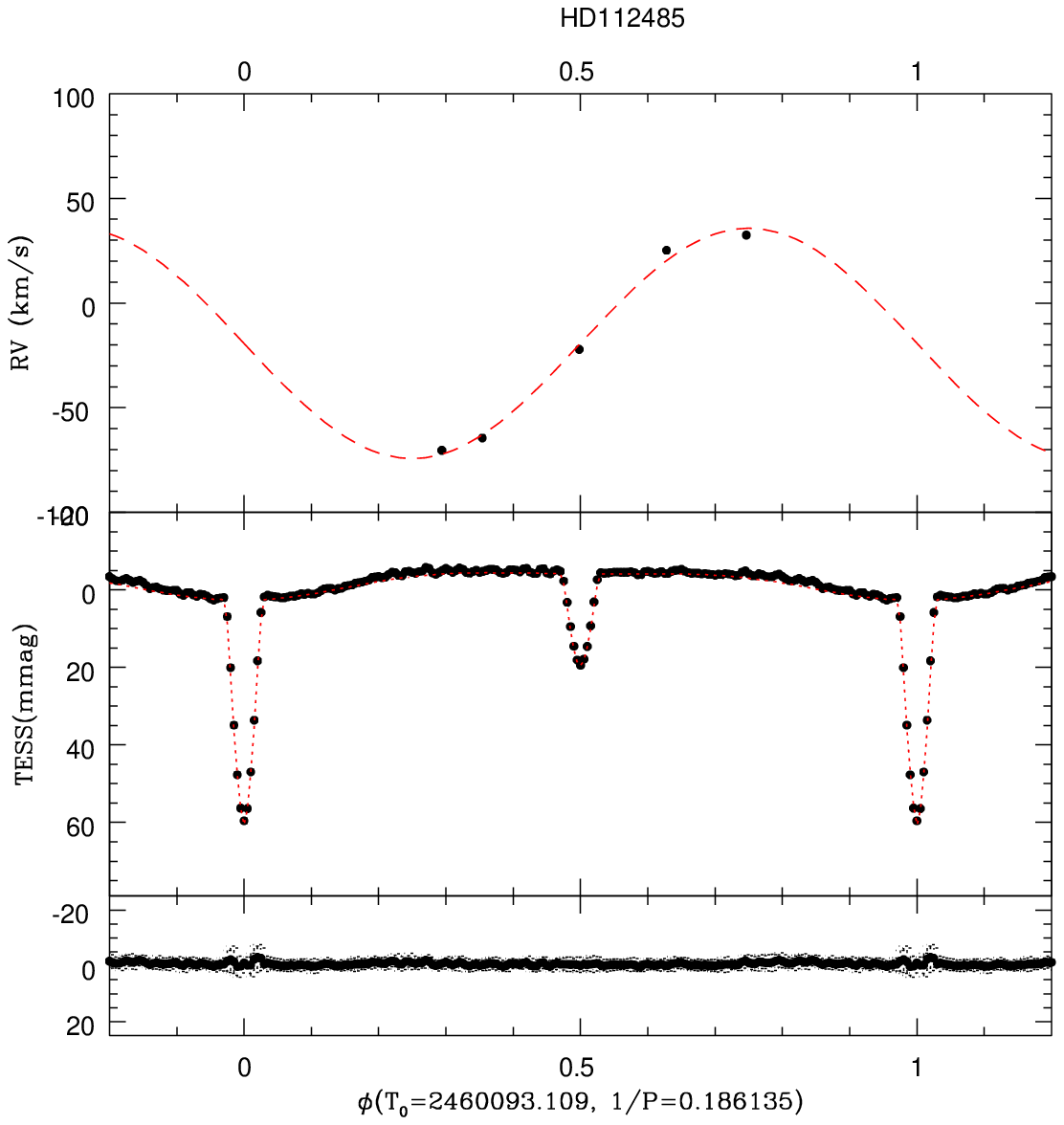}
    \includegraphics[width=5.5cm]{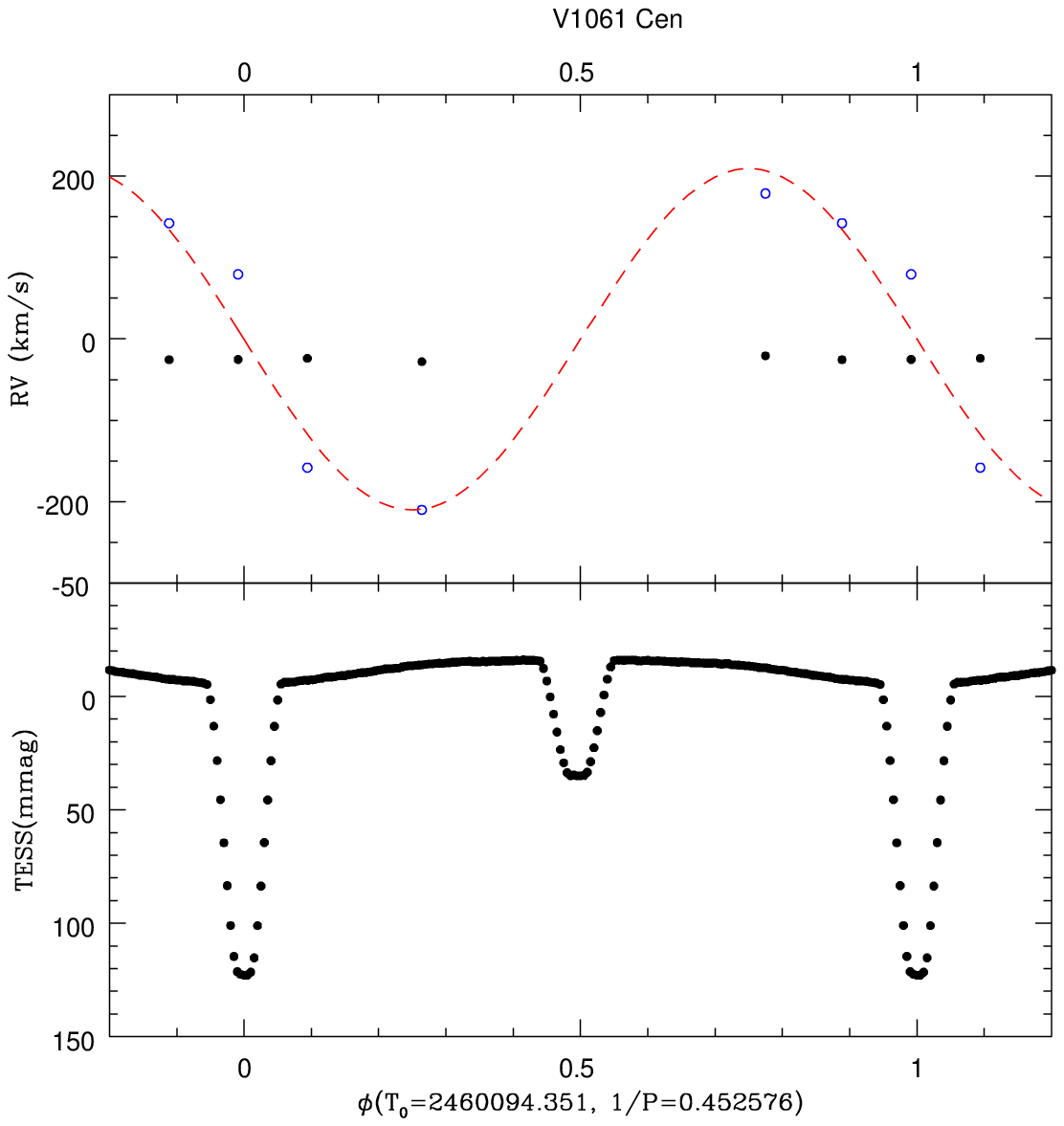}
    \includegraphics[width=5.5cm]{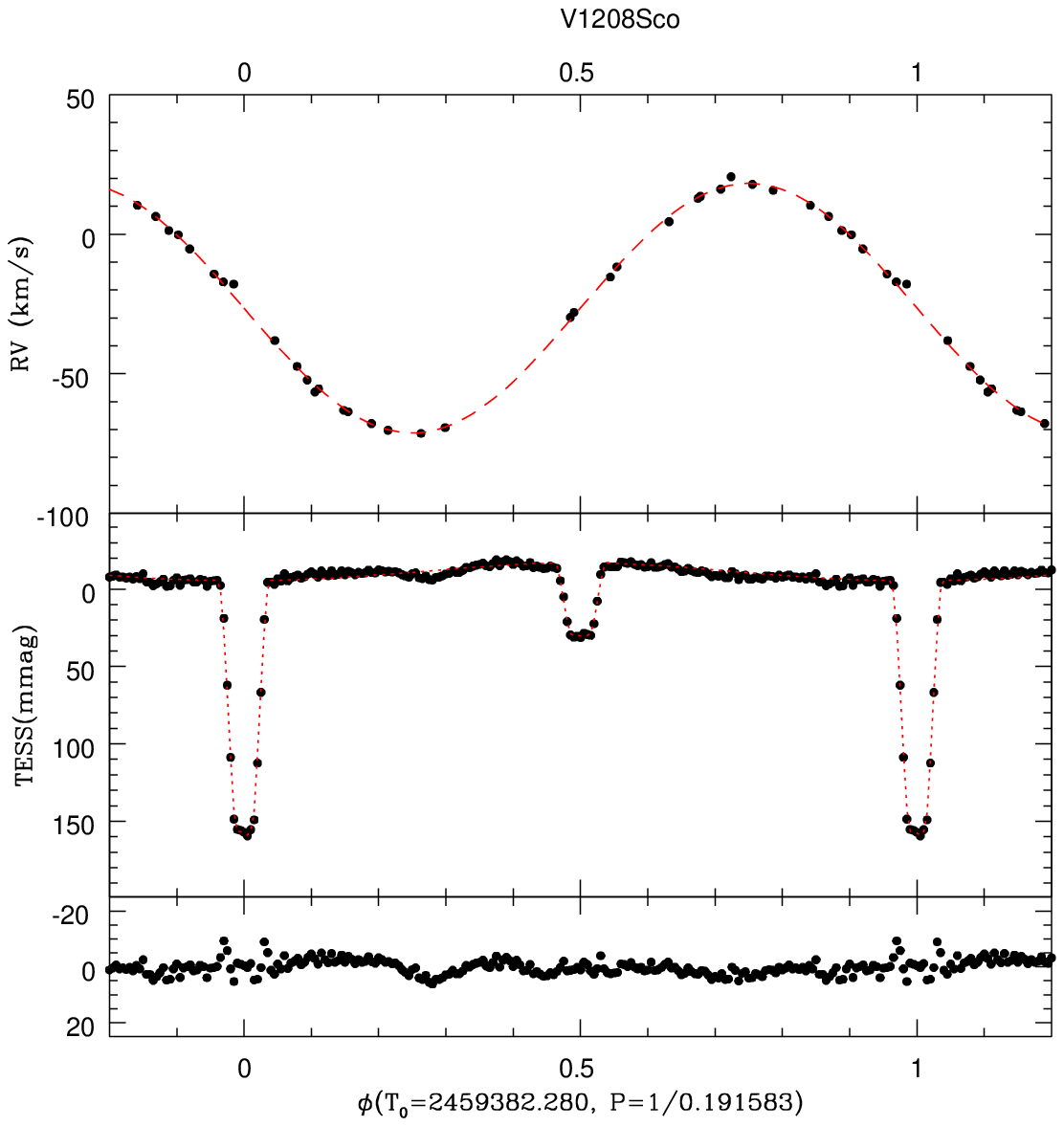}
    \includegraphics[width=5.5cm]{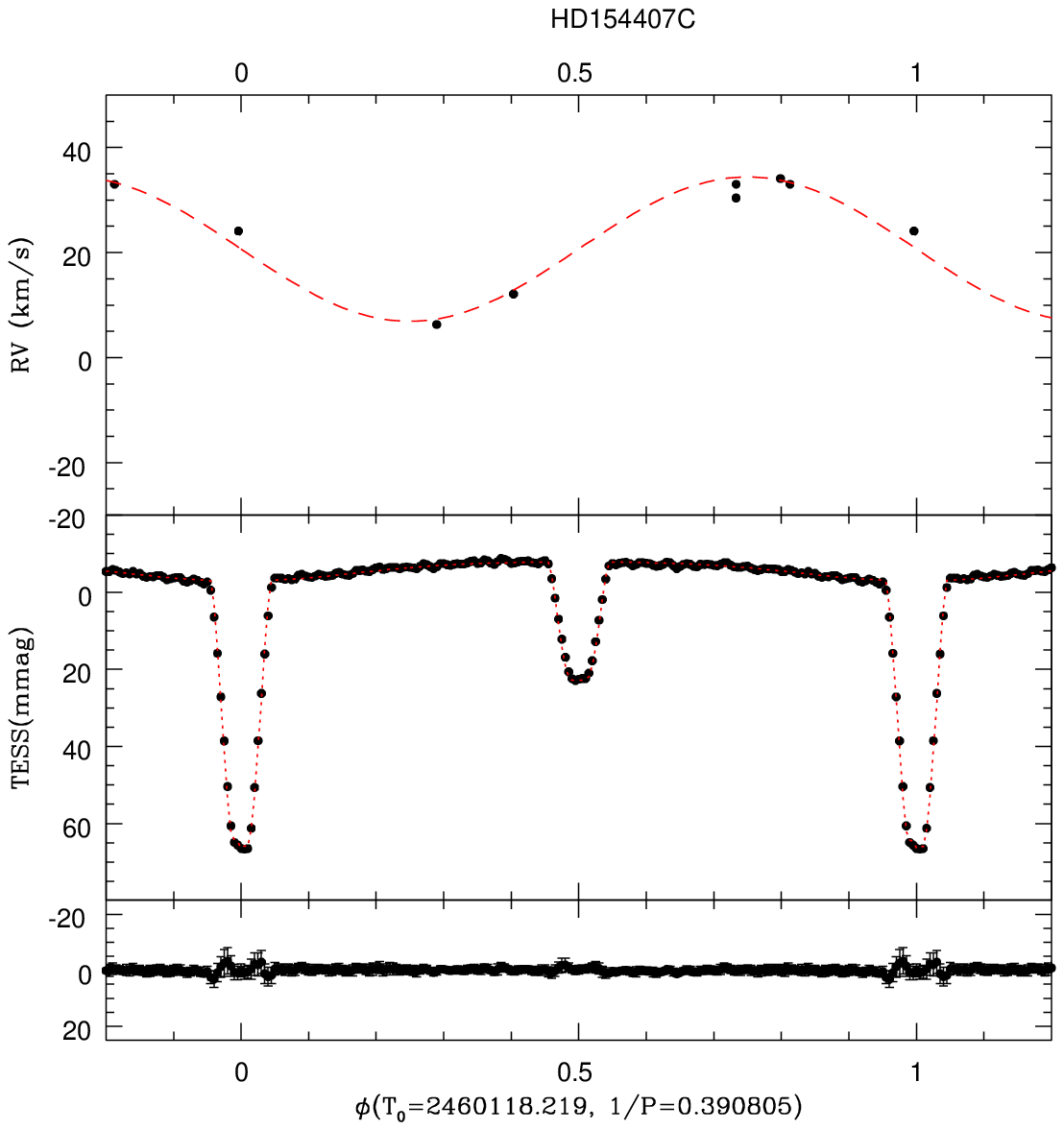}
    \includegraphics[width=5.5cm]{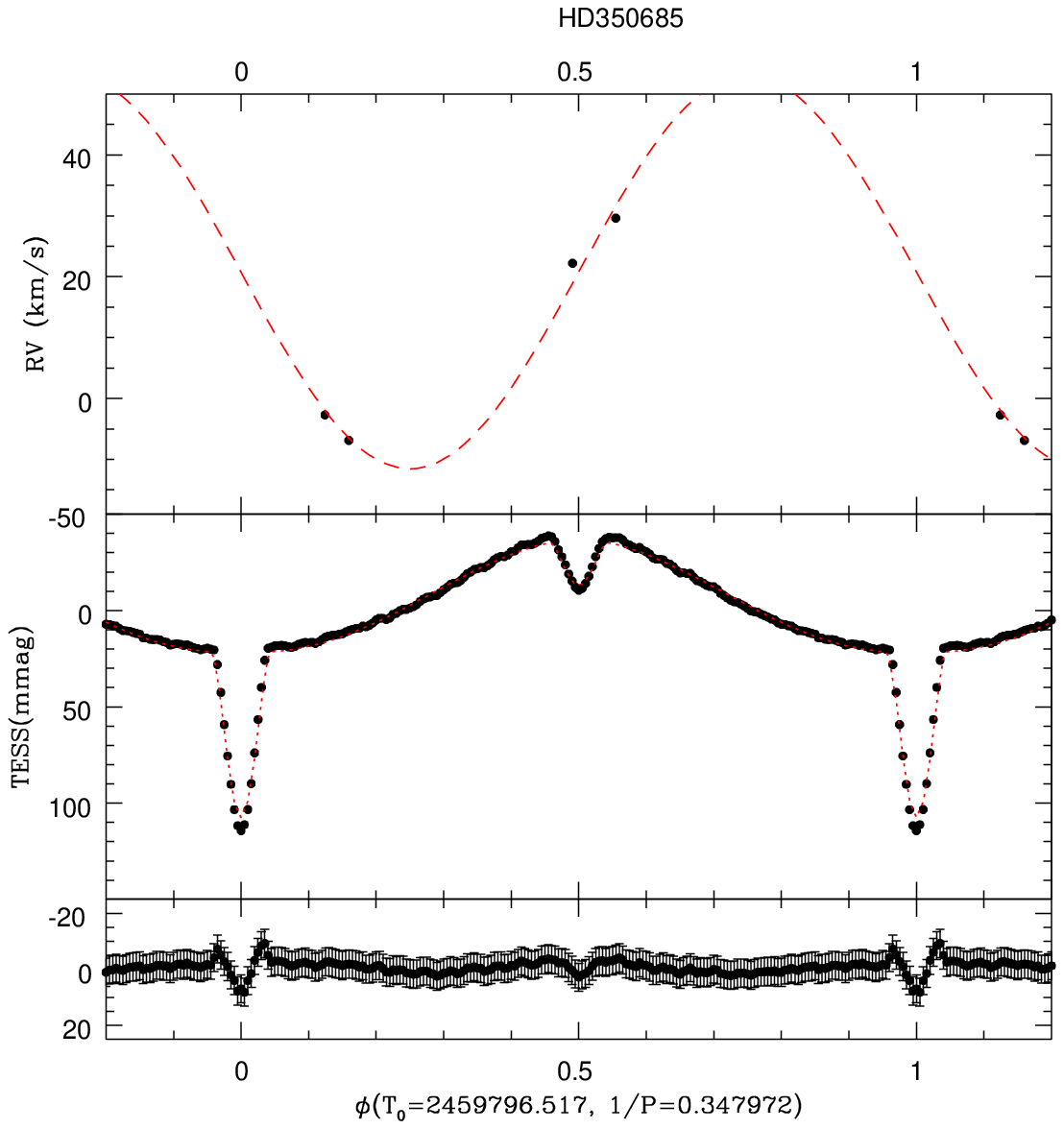}
  \end{center}  
  \caption{Binned and averaged light curves of all targets (bottom panels - the data used to produce these curves are identified by boldface in Table \ref{targetlist}; residuals are shown if a fitting was achieved), together with RV curves (top panels). The folding was done using the ephemerides of Table \ref{phot1}. The best-fit solutions from Table \ref{binsol} are superimposed to the data using red lines. For LS\,I\,+61\,275, the black RV points come from APOGEE while the red ones come from \citet{hua06b}, showing the discrepancy between them. For V1061\,Cen, the black filled dots and blue circles correspond to the main and fainter spectral components, respectively (see Table \ref{rv} and text for details). \label{rvphot}}
\end{figure*}

A sample of individual light curves of each target is shown in Fig. \ref{LC} (the full set is available on Zenodo), while folded ones are shown in Fig. \ref{rvphot}, along with the folded RV curves. Literature velocities are recalled in Table \ref{litrv} and our newly determined RVs are provided in Table \ref{rv}. Table \ref{whole} lists the whole sample of nascent, short-period massive binaries with extreme mass ratios and Fig. \ref{histo} provides histograms of the rotational velocities, highlighting their high values.

Finally, the photometric data were cleaned from the binary signals using the average folded and binned light curves. A period search was then performed on the residuals. A forest of low-frequency signals with increasing amplitudes (``red noise'') are obvious in these frequency spectra, notably for LS\,I\,+61\,145, TYC\,741-1565-1, HD\,154407C, and V1208\,Sco. While we cannot fully exclude the presence of imperfect calibration effects and some small long-term variations in these curves, stochastic low-frequency signals are well known to be present in massive stars and our targets would thus simply be no exception. In addition, isolated signals are also present. Those with low frequencies ($\nu<4$\,d$^{-1}$) usually indicate Slowly Pulsating B-stars (SPB) activity. In our sample, low-frequency signals are detected for AN\,Dor, HD\,254346, HD\,309317, HD\,112485, and V1061\,Cen. In contrast, isolated signals at high frequencies ($\nu>4$\,d$^{-1}$) usually indicate the presence of $\beta$\,Cep pulsations. \citet{eze24} already reported the detection of $\beta$\,Cep activity in LS\,I\,+61\,145, HD\,112485, HD\,254346, and V1061\,Cen, plus AN\,Dor (see also \citealt{sou22}) and HD\,154407C as candidates. We confirm their detections and add to the list the cases of HD\,309317, HD\,350685, and V1208\,Sco, plus TYC\,1881-933-1 as candidate. The cases of HD\,154407C and V1208\,Sco must of course be taken with caution as the \te\ data alone cannot disentangle the source of pulsations which could thus be these stars or their close neighbors. Nevertheless, the presence of both eclipsing and pulsational signals provides a golden opportunity for asteroseismology. The orbital analysis helps pinpoint the stellar properties, which can then serve as input for a detailed asteroseismic modeling. The fact that most of our sources are rotating quite fast constitutes both a challenge and an asset in this domain, as the pulsations of such stars remain difficult to model. 

\FloatBarrier

\begin{table}
  \scriptsize
  \caption{RVs found in literature (\citealt{hua06} - first line - and \citealt{jon20} for LS\,I\,+61\,275, \citealt{zha22} for TYC\,1881-933-1 and \citealt{ban22} for V1208\,Sco, after correcting some typos in the latter).
 \label{litrv}}
  \begin{tabular}{lcccc}
    \hline
Name & $HJD-2\,450\,000$ & RV (\kms) & $HJD-2\,450\,000$ & RV (\kms)\\
\hline
LS\,I\,+61\,275 & 1861.825 &  --1.1 & 1863.820 & --98.2 \\
LS\,I\,+61\,275 & 7650.873 &    5.6$\pm$0.2 & 7764.605 & --11.6$\pm$0.9 \\
LS\,I\,+61\,275 & 7677.805 &  --0.0$\pm$0.3 & 7766.632 & --24.0$\pm$0.4 \\
LS\,I\,+61\,275 & 7710.764 &  --9.0$\pm$0.4 & 8008.986 &    4.4$\pm$0.4 \\
LS\,I\,+61\,275 & 7762.627 & --27.2$\pm$0.2 & 8037.888 &    1.0$\pm$0.3 \\
TYC\,1881-933-1 & 8084.240 &  --3.7$\pm$1.3 & 8122.161 &   31.6$\pm$6.1  \\
TYC\,1881-933-1 & 8084.272 &    6.3$\pm$1.5 & 8122.222 &   28.6$\pm$3.9  \\
TYC\,1881-933-1 & 8086.196 & --14.0$\pm$1.4 & 8179.987 & --15.2$\pm$3.9  \\
TYC\,1881-933-1 & 8086.213 & --16.9$\pm$1.9 & 8180.010 &  --9.8$\pm$5.8  \\
TYC\,1881-933-1 & 8086.246 &    0.0$\pm$3.0 & 8180.027 &  --6.1$\pm$6.1  \\
TYC\,1881-933-1 & 8086.263 &  --5.4$\pm$1.4 & 8447.258 &  --3.7$\pm$14.2 \\
TYC\,1881-933-1 & 8086.280 &    6.2$\pm$2.0 & 8447.275 &   24.0$\pm$8.6  \\
TYC\,1881-933-1 & 8086.296 &    8.3$\pm$2.7 & 8447.291 &  --8.3$\pm$18.9 \\
TYC\,1881-933-1 & 8088.183 & --20.0$\pm$2.6 & 8447.307 &    9.2$\pm$21.0 \\
TYC\,1881-933-1 & 8088.237 & --14.2$\pm$3.1 & 8447.323 &  --2.1$\pm$9.6  \\
TYC\,1881-933-1 & 8088.253 & --13.9$\pm$1.7 & 8558.998 &    7.3$\pm$11.7 \\
TYC\,1881-933-1 & 8089.197 &  --1.6$\pm$7.2 & 8559.014 &  --3.1$\pm$7.1  \\
TYC\,1881-933-1 & 8089.206 &  --0.4$\pm$6.3 & 8801.275 &   39.7$\pm$1.0  \\
TYC\,1881-933-1 & 8089.216 & --13.0$\pm$7.3 & 8801.291 &   47.8$\pm$2.1  \\
TYC\,1881-933-1 & 8090.189 &  --1.9$\pm$4.9 & 8801.307 &   46.3$\pm$2.0  \\
TYC\,1881-933-1 & 8090.198 & --15.9$\pm$6.6 & 8862.091 & --20.2$\pm$3.4  \\
TYC\,1881-933-1 & 8090.208 & --18.4$\pm$3.2 & 8862.107 & --20.8$\pm$3.5  \\
TYC\,1881-933-1 & 8114.117 &   55.4$\pm$1.3 & 8862.123 & --19.6$\pm$5.2  \\
TYC\,1881-933-1 & 8114.154 &   56.7$\pm$1.8 & 8891.052 &   34.6$\pm$4.2  \\
TYC\,1881-933-1 & 8114.170 &   52.8$\pm$2.3 & 8891.069 &   46.7$\pm$5.5  \\
TYC\,1881-933-1 & 8114.198 &   55.0$\pm$1.4 & 8891.085 &   45.2$\pm$4.5  \\
TYC\,1881-933-1 & 8122.144 &   28.2$\pm$4.4 \\
V1208\,Sco & 7845.774 &    4.5$\pm$0.2 & 8001.599 & --29.8$\pm$0.2 \\
V1208\,Sco & 7919.658 &   15.7$\pm$0.2 & 8001.627 & --28.0$\pm$0.2 \\
V1208\,Sco & 7920.693 & --17.9$\pm$0.1 & 8003.602 &    6.4$\pm$0.2 \\
V1208\,Sco & 7926.542 & --56.5$\pm$0.1 & 8004.526 & --38.1$\pm$0.1 \\
V1208\,Sco & 7929.691 &   16.2$\pm$0.2 & 8014.494 & --14.3$\pm$0.1 \\
V1208\,Sco & 7932.586 & --71.4$\pm$0.2 & 8015.533 & --63.6$\pm$0.1 \\
V1208\,Sco & 7934.733 &   12.8$\pm$0.1 & 8017.567 & --15.3$\pm$0.1 \\
V1208\,Sco & 7946.710 & --17.1$\pm$0.2 & 8019.525 &  --5.3$\pm$0.2 \\
V1208\,Sco & 7947.644 & --63.1$\pm$0.1 & 8020.525 & --55.4$\pm$0.1 \\
V1208\,Sco & 7950.649 &   20.6$\pm$0.2 & 8021.506 & --69.3$\pm$0.1 \\
V1208\,Sco & 7955.627 &   13.5$\pm$0.1 & 8221.835 &   13.5$\pm$0.1 \\
V1208\,Sco & 7957.721 & --47.4$\pm$0.1 & 8236.846 & --11.7$\pm$0.1 \\
V1208\,Sco & 7963.520 & --67.9$\pm$0.1 & 8250.729 & --70.2$\pm$0.2 \\
V1208\,Sco & 7971.693 &   17.9$\pm$0.1 & 8264.688 &    1.3$\pm$0.2 \\
V1208\,Sco & 7982.582 &   10.4$\pm$0.1 & 8286.640 & --52.2$\pm$0.1 \\
V1208\,Sco & 7998.559 &  --0.2$\pm$0.1 \\
\hline
  \end{tabular}
\end{table}

\begin{table}
  \scriptsize
  \caption{RVs derived for all targets from new spectra.
 \label{rv}}
  \begin{tabular}{lcc}
    \hline
Name & $HJD-2\,450\,000$ & RV (\kms) \\
\hline
LS\,I\,+61\,145 & 10660.323 & $-44.7\pm1.2$ \\
LS\,I\,+61\,145 & 10660.526 & $-56.0\pm0.8$ \\
AN\,Dor         &  5187.536 & $100.7\pm4.0$ \\ 
AN\,Dor         &  5190.670 & $ -0.8\pm4.3$ \\
AN\,Dor         &  5541.529 & $ 83.7\pm3.9$ \\
AN\,Dor         &  5541.538 & $ 83.2\pm3.9$ \\
AN\,Dor         &  5541.732 & $ 41.6\pm4.2$ \\
AN\,Dor         &  5541.841 & $ 23.4\pm4.1$ \\
AN\,Dor         &  8830.543 & $ 36.4\pm3.2$ \\
AN\,Dor         &  8840.565 & $ 75.7\pm3.1$ \\
AN\,Dor         &  8888.560 & $ 71.4\pm3.2$ \\
AN\,Dor         & 10588.860 & $ 38.5\pm4.0$ \\
AN\,Dor         & 10589.872 & $ 66.6\pm3.9$ \\
AN\,Dor         & 10672.656 & $  0.6\pm4.2$ \\
AN\,Dor         & 10673.576 & $114.2\pm3.9$ \\
AN\,Dor         & 10687.649 & $102.7\pm4.0$ \\
HD\,254346 	& 10290.527 & $ 13.2\pm4.9$ \\
HD\,254346 	& 10292.531 & $ 22.5\pm4.8$ \\
HD\,254346 	& 10293.546 & $  6.9\pm4.7$ \\
HD\,254346 	& 10294.580 & $ -4.4\pm4.7$ \\
TYC\,741-1565-1 & 10652.847 & $ 40.9\pm4.4$ \\
TYC\,741-1565-1 & 10658.590 & $ 53.3\pm4.2$ \\
TYC\,741-1565-1 & 10658.709 & $ 49.5\pm4.5$ \\
TYC\,741-1565-1 & 10685.732 & $ 51.4\pm4.2$ \\
TYC\,741-1565-1 & 10696.678 & $ 44.3\pm4.2$ \\
TYC\,741-1565-1 & 10727.659 & $ 35.4\pm4.3$ \\
TYC\,741-1565-1 & 10729.677 & $ 37.4\pm4.3$ \\
HD\,309317 	&  9238.639 & $-30.7\pm4.0$ \\
HD\,309317 	&  9238.842 & $-36.9\pm3.8$ \\
HD\,309317 	&  9239.643 & $ -1.7\pm4.1$ \\
HD\,309317 	&  9239.855 & $ 12.4\pm4.0$ \\
HD\,309317 	&  9241.632 & $-12.2\pm3.9$ \\
HD\,309317 	&  9241.860 & $ -0.7\pm3.9$ \\
HD\,309317 	&  9296.765 & $  6.1\pm3.9$ \\
HD\,309317 	&  9297.788 & $-32.7\pm3.9$ \\
HD\,112485 	& 10649.843 & $ 25.2\pm4.9$ \\
HD\,112485 	& 10701.773 & $-70.4\pm4.7$ \\
HD\,112485 	& 10702.871 & $-22.1\pm4.9$ \\
HD\,112485 	& 10712.842 & $-64.5\pm4.8$ \\
HD\,112485 	& 10725.695 & $ 32.5\pm4.9$ \\
V1061\,Cen 	& 10725.791 & $ 178.5\pm11.7 (-20.8\pm1.6)$ \\
V1061\,Cen 	& 10728.706 & $-158.2\pm14.8 (-23.9\pm1.9)$ \\
V1061\,Cen 	& 10730.689 & $  79.3\pm12.6 (-25.2\pm0.9)$ \\
V1061\,Cen 	& 10734.881 & $ 142.0\pm18.3 (-25.6\pm1.8)$ \\
V1061\,Cen 	& 10735.711 & $-210.1\pm 3.6 (-28.0\pm0.8)$ \\
HD\,154407C 	& 10587.515 & $ 12.1\pm4.5$ \\
HD\,154407C 	& 10588.527 & $ 34.1\pm4.3$ \\
HD\,154407C 	& 10588.563 & $ 33.0\pm4.3$ \\
HD\,154407C 	& 10734.885 & $ 24.1\pm4.5$ \\
HD\,154407C 	& 10741.888 & $ 33.0\pm4.5$ \\
HD\,154407C 	& 10745.871 & $  6.3\pm4.7$ \\
HD\,154407C 	& 10777.711 & $ 30.4\pm4.5$ \\
HD\,350685 	& 10440.604 & $ -2.7\pm5.2$ \\
HD\,350685 	& 10441.658 & $ 22.2\pm5.1$ \\
HD\,350685 	& 10443.579 & $ -6.9\pm5.1$ \\
HD\,350685 	& 10444.715 & $ 29.6\pm5.0$ \\
\hline
  \end{tabular}
\tablefoot{SOPHIE, FEROS, and Hermes spectra include the heliocentric correction while XShooter or CARMENES spectra do not but the RVs quoted above are all corrected. The RVs of the stationary component of V1061\,Cen are provided between parentheses. }
\end{table}

\begin{figure*}
  \begin{center}
    \includegraphics[width=8cm]{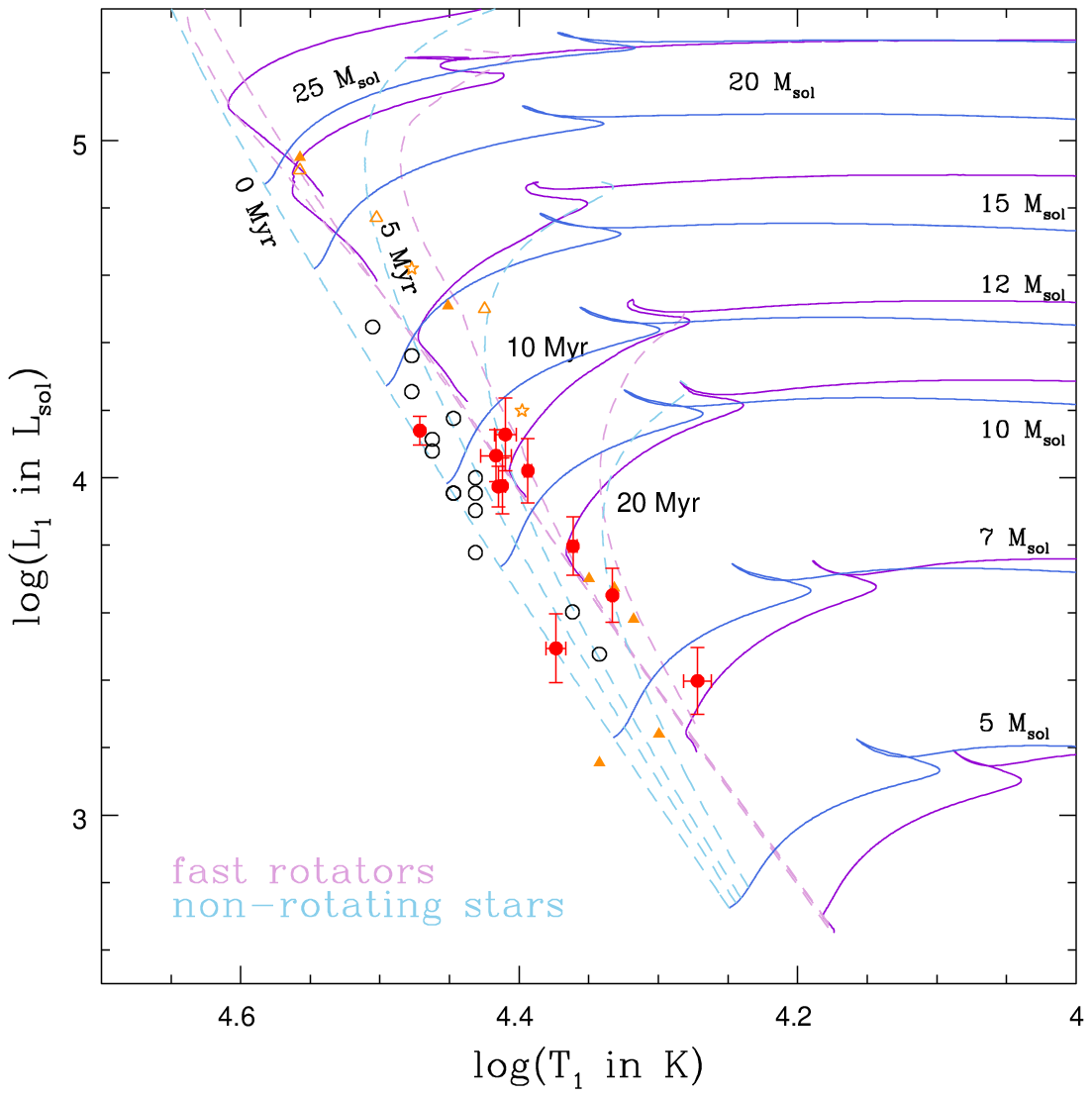}
    \includegraphics[width=8cm]{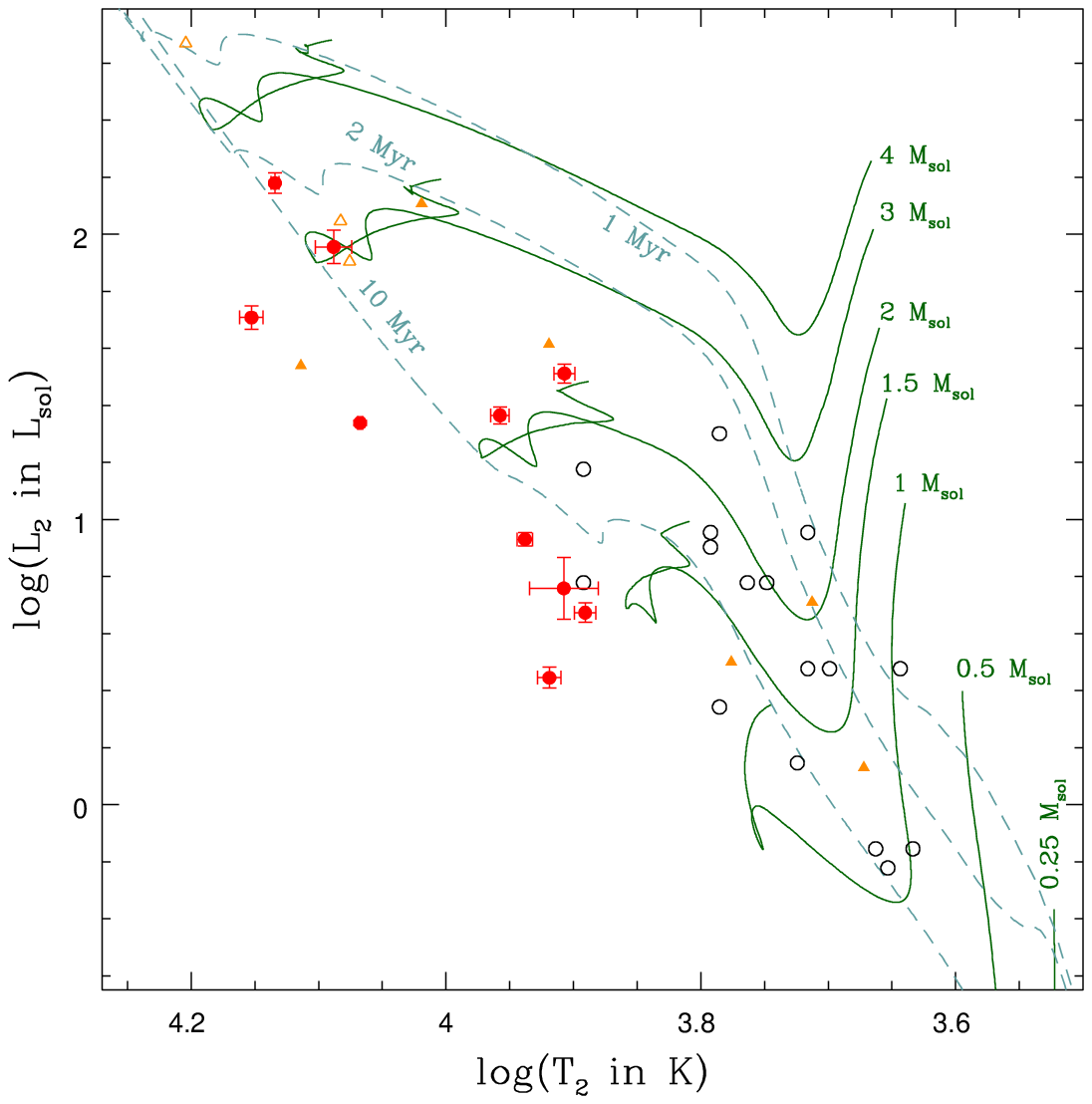}
    \includegraphics[width=8cm]{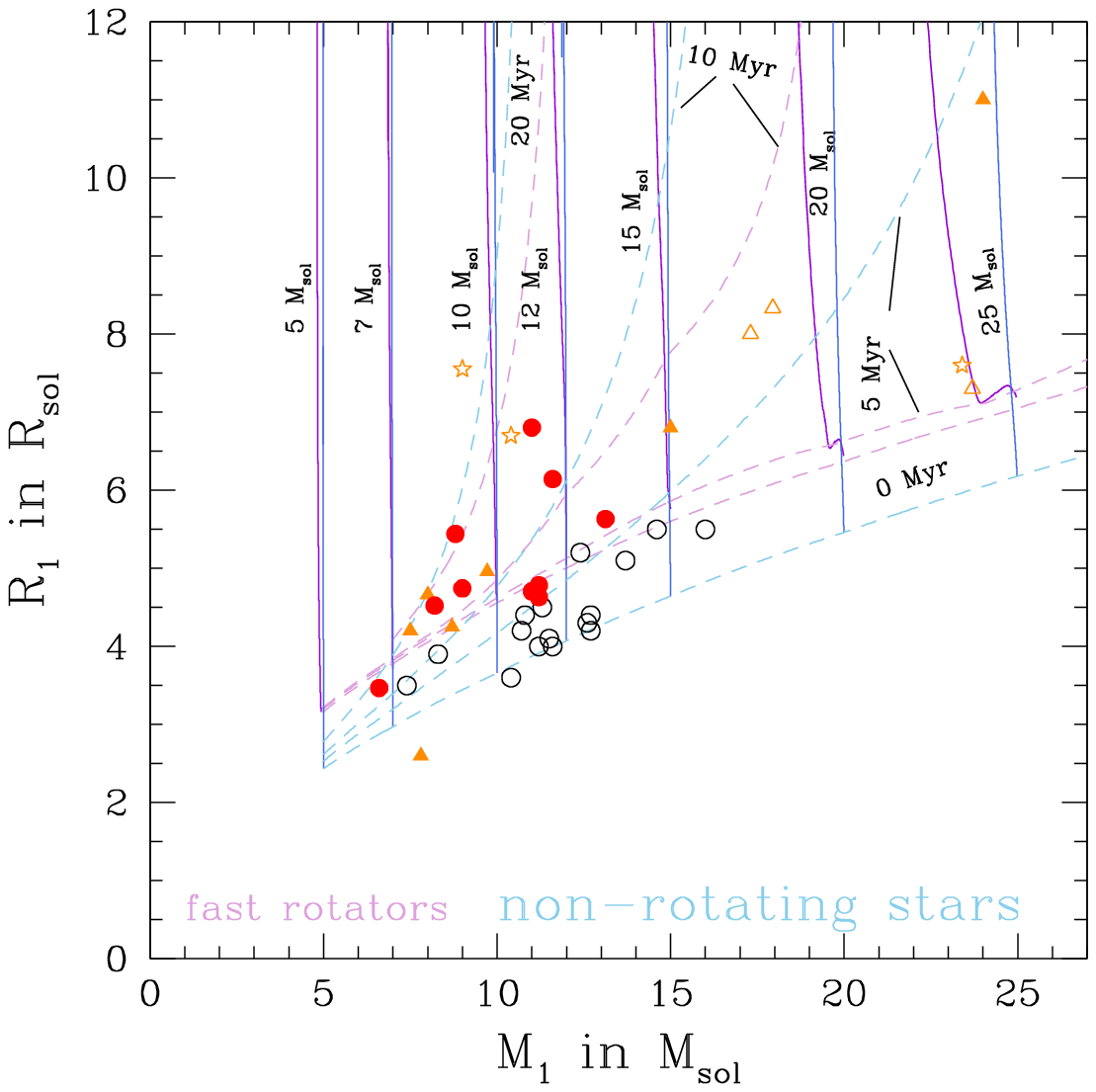}
    \includegraphics[width=8cm]{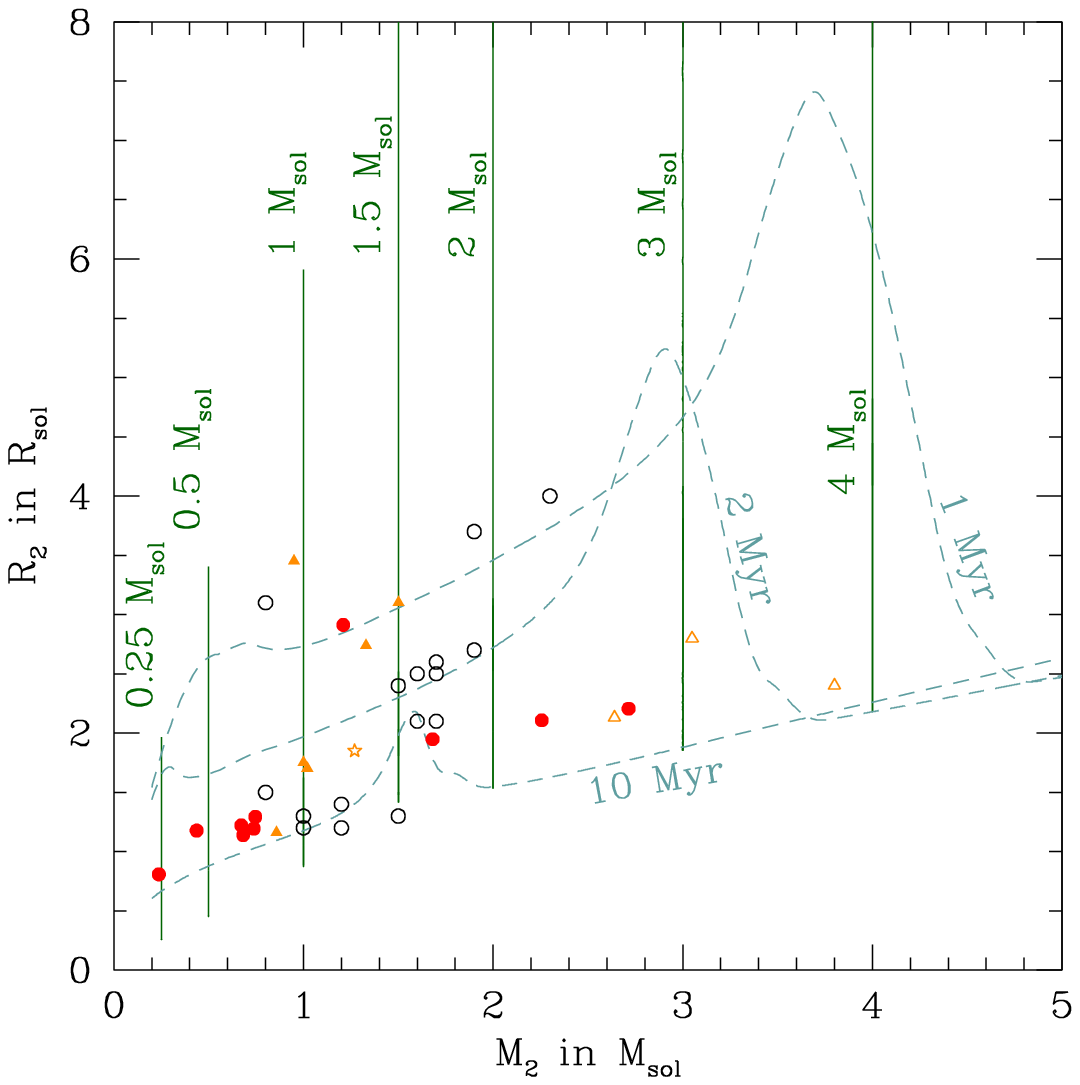}
  \end{center}  
  \caption{Hertzsprung-Russell diagrams for primaries (top left, with the post-PMS evolutionary tracks of \citet{bro11} without initial rotation in blue and with the fastest initial rotation available in these models in violet lines) and secondaries (top right, with the PMS tracks of \citealt{tog11} in green). Isochrones are shown in lighter color and with dashed lines. Symbols as in Fig. \ref{agem}. For our targets, the secondary effective temperatures are those without reflection and the bolometric luminosities were calculated from these temperatures and the stellar radii (Table \ref{binsol}) while the bolometric luminosities of the primaries are those from Table \ref{targetlist}. Bottom panels provide the mass-radius diagrams, using the same conventions. The comparison with theoretical evolutionary tracks is not without limitations: we recall that the errors shown here are most probably underestimated (see Sect. 3.3) and that the tracks correspond to single stars, not to binary systems with strong reflection effects.  \label{hrd}}
\end{figure*}

\begin{table}
  \scriptsize
  \caption{The whole Galactic sample of nascent short-period ($P<10$\,d) binaries with early-type primaries and extreme mass ratios (sp. type $\le$B3 and $q<0.2$, see text for details).
 \label{whole}}
  \begin{tabular}{lccccccl}
    \hline
Name & Sp. Type & $P$ & $q$ & $e$ & $v \sin(i)$ & Age & Ref\\
     &          & (d) &     &     &  (\kms)     & (Myr) & \\  
\hline
LS\,I+61\,145   &B1V    &1.85 &0.06 &0    &110$^*$ &4.9  & a \\ 
HD\,25631       &B3V    &5.24 &0.13 &0    &221     &9.6  & b \\
TYC\,1881-933-1 &B2V    &1.02 &0.11 &0    &143$^*$ &29.3 & a \\ 
HD\,254346      &B1.5V  &5.43 &0.06 &0.015&201     &11.6 & a \\ 
HD\,46485       &O7V    &6.94 &0.04 &0.03 &334     &2.5  & b \\
TYC\,741-1565-1 &B2V    &2.26 &0.03 &0    &149     &3.9  & a \\ 
HD\,62747       &B1.5III&3.93 &0.14 &0    &100$^*$ &16.2 & c,d \\
HD\,309317      &B2V    &2.26 &0.08 &0    &203     &11.8 & a \\ 
HD\,120307      &B2IV   &2.63 &0.12 &0    &65$^*$  &11.1 & e \\
HD\,138690A     &B2IV   &2.85 &0.17 &0    &236     &16.7 & e \\
HD\,145482      &B2.5V  &5.77 &0.15 &0.19 &170     &8.0  & c,d,f,g \\
HD\,149834      &B2V    &4.60 &0.09 &0.04 &216     &5.9  & h \\
HD\,152200      &B0V    &4.44 &     &0.04 &240     &4.3  & i \\
V1208\,Sco      &B0.5V  &5.22 &0.17 &0    &87      &0.6  & a \\ 
HD\,154407C     &B1V    &2.56 &0.04 &0    &145     &4.8  & a \\ 
HD\,158926A     &B1.5IV &5.95 &0.17 &0.26 &125     &11   & j  \\
HD\,350685      &B1.5V  &2.87 &0.11 &0    &248     &6.2  & a \\ 
HD\,191495      &B0V    &3.64 &0.10 &0    &201     &7.6  & b \\
\hline
HD\,163892      &O9.5IV &7.84 &0.18 &0.04 &212     &     & k \\
HD\,165246      &O8V    &4.59 &0.16 &0.03 &253     &3.3  & l \\
HD\,210478      &B1V    &3.81 &0.15 &0    &120$^*$ &10   & m \\
\hline
  \end{tabular}
\tablefoot{References: (a) this work, (b) \citet{naz23}, (c) \citet{pig24}, (d) \citet{sim17}, (e) \citet{jer21}, (f) \citet{lev87}, (g) \citet{gul16}, (h) \citet{sta21}, (i) \citet{ban23} - the quoted age is the cluster age, (j) \citet{tan06} and references therein, (k) \citet{mah22}, (l) \citet{joh21}, and (m) \citet{car14}. Asterisks $*$ in the sixth column indicate the synchronized systems. The last three objects may be ZAMS systems rather than nascent ones (see text for details).}
\end{table}

\begin{figure}
  \begin{center}
    \includegraphics[width=8cm]{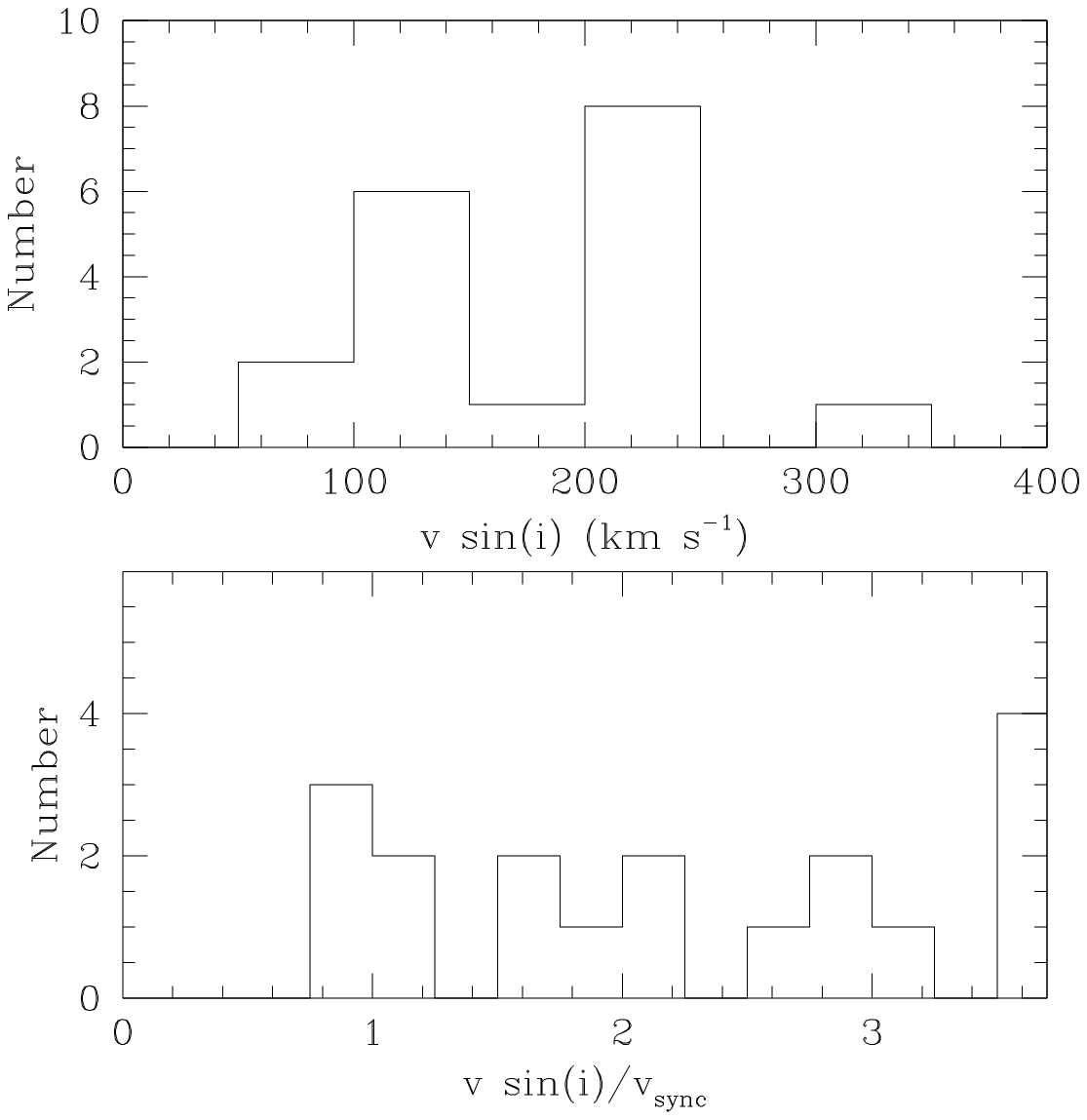}
  \end{center}  
  \caption{Histogram of the projected rotational velocities $v \sin(i)$ of the 18 nascent short-period binaries with extreme mass ratios (Table \ref{whole}) and of the ratios between these velocities to the synchronization velocities (last bin comprises all $v \sin(i)/v_{sync}>3.5$ cases).  \label{histo}}
\end{figure}

\end{document}